\documentclass[aps,pra,twocolumn,showpacs,letterpaper]{revtex4}

\usepackage{amsmath}
\usepackage{bbm}
\usepackage{amsfonts}
\usepackage{graphicx}

\def\Ecore{{\cal E}}
\def\Eqn#1{{Eq.~(\ref{#1})}}
\def\Eqns#1#2{{Eqs.~(\ref{#1},~\ref{#2})}}
\def\Eqnsd#1#2{{Eqs.~(\ref{#1}-\ref{#2})}}

\def\figdir{{./}}

\begin{document}
\title{Operation of a superconducting nanowire quantum interference device \\
with mesoscopic leads} 
\date{\today}
  
\begin{abstract}
A theory describing the operation of a superconducting nanowire
quantum interference device (NQUID) is presented.  The device consists
of a pair of thin-film superconducting leads connected by a pair of
topologically parallel ultra-narrow superconducting wires.  It
exhibits intrinsic electrical resistance, due to thermally-activated
dissipative fluctuations of the superconducting order parameter.
Attention is given to the dependence of this resistance on the
strength of an externally applied magnetic field aligned perpendicular
to the leads, for lead dimensions such that there is essentially
complete and uniform penetration of the leads by the magnetic field.
This regime, in which at least one of the lead dimensions---length or
width---lies between the superconducting coherence and penetration
lengths, is referred to as the {\it mesoscopic\/} regime.  The
magnetic field causes a pronounced oscillation of the device
resistance, with a period {\it not\/} dominated by the Aharonov-Bohm
effect through the area enclosed by the wires and the film edges but,
rather, in terms of the geometry of the leads, in contrast to the
well-known Little-Parks resistance of thin-walled superconducting
cylinders.  A detailed theory, encompassing this phenomenology
quantitatively, is developed through extensions, to the setting of
parallel superconducting wires, of the
Ivanchenko-Zil'berman-Ambegaokar-Halperin theory of intrinsic
resistive fluctuations in a current-biased Josephson junctions and the
Langer-Ambegaokar-McCumber-Halperin theory of intrinsic resistive
fluctuations in superconducting wires.  In particular, it is
demonstrated that via the resistance of the NQUID, the wires act as a
probe of spatial variations in the superconducting order parameter
along the perimeter of each lead: in essence, a superconducting phase
gradiometer.
\end{abstract}

\author{David Pekker, Alexey Bezryadin, David S. Hopkins, and Paul
M.~Goldbart}

\affiliation{Department of Physics, University of Illinois at
Urbana-Champaign, 1110 West Green Street, Urbana, Illinois 61801-3080,
U.S.A.  }


\maketitle

\section{Introduction}
The Little-Parks effect concerns the electrical resistance of a thin
cylindrically-shaped superconducting film and, specifically, the
dependence of this resistance on the magnetic flux threading the
cylinder~\cite{LP,Tinkham}. It is found that the resistance is a
periodic function of the magnetic field, with period inversely
proportional to the cross-sectional area of the cylinder. Similarly,
in a DC SQUID, the critical value of the supercurrent is periodic in
magnetic field, with period inversely proportional to the area
enclosed by the SQUID ring~\cite{Tinkham}.

In this Paper, we consider a mesoscopic analog of a DC SQUID.  The
analog consists of a device composed of a thin superconducting film
patterned into two mesoscopic leads that are connected by a pair of
(topologically) parallel, short, weak, superconducting wires. Thus, we
refer to the device as an NQUID (superconducting nanowire quantum
interference device).  The only restriction that we place on the wires
of the device is that they be thin enough for the order parameter to
be taken as constant over each cross-section of a wire, varying only
along the wire length.  In principle, this condition of
one-dimensionality is satisfied if the wire is much thinner than the
superconducting coherence length $\xi$.  In practice, it is
approximately satisfied provided the wire diameter $d$ is smaller than
$4.4\,\xi$ ~\cite{LikharevRMP}. For thicker wires, vortices can exist
inside the wires, and such wires may not be assumed to be one
dimensional.

By the term {\it mesoscopic\/} we are characterizing phenomena that
occur on length-scales larger than the superconducting coherence
length \(\xi\) but smaller than the electromagnetic penetration depth
\(\lambda_\perp\) associated with magnetic fields applied
perpendicular to the superconducting film.  We shall call a lead
mesoscopic if at least one of its two long dimensions is in the
mesoscopic regime; the other dimension may be either mesoscopic or
macroscopic. Thus, a weak magnetic field applied perpendicular to a
mesoscopic lead will penetrate the lead without appreciable
attenuation and without driving the lead from the homogeneous
superconducting state to the Abrikosov vortex state. This is similar
to the regime of operation of superconducting wire networks; see e.g.,
Ref.~\cite{AB1}.  The nanowires connecting the two leads are taken to
be topologically parallel (i.e.~parallel in the sense of electrical
circuitry): these nanowires and edges of the leads define a closed
geometrical contour, which will be referred to as the {\it
Aharonov-Bohm (AB) contour}. In our approach, the nanowires are
considered to be links sufficiently weak that any effects of the
nanowires on the superconductivity in the leads can be safely ignored.

\begin{figure}
\includegraphics[width=8cm]{\figdir/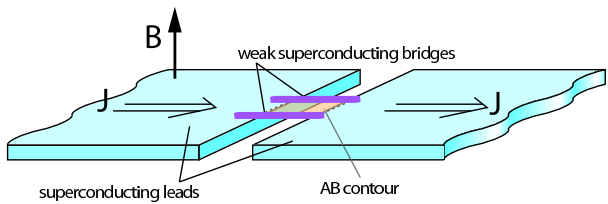}
\caption{\label{device} Schematic depiction of the superconducting
phase gradiometer. A current \(J\) is passed through the bridges in
the presence of a perpendicular magnetic field of strength \(B\). }
\end{figure}

The theory presented here has been developed to explain experiments
conducted on DNA-templated NQUIDs~\cite{us_in_science}. These
experiments measure the electrical resistivity of a pair of
superconducting nanowires suspended between long superconducting
strips (see Fig.~\ref{device}). In them, a current source is used to
pass DC current from a contact on the far end of the left lead to one
on the far end of the right lead. The voltage between the contacts is
measured (and the resistance is hence determined) as a function of the
magnetic field applied perpendicular to the plane of the strips.

\begin{figure*}
\includegraphics[width=15cm]{\figdir/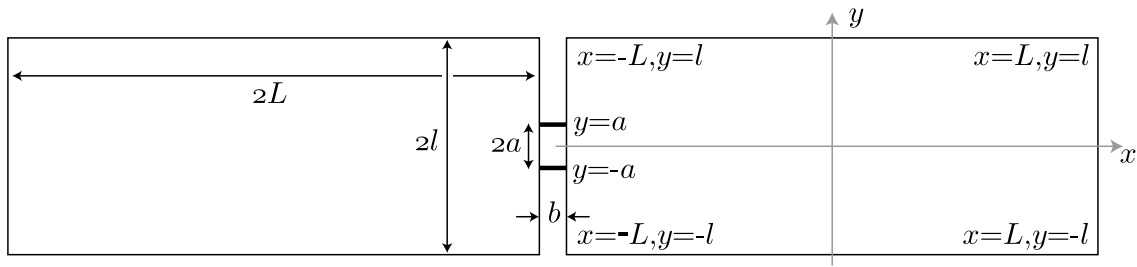}
\caption{Geometry of the two-wire device, showing the dimensions. The
coordinate system used for the right lead (with the origin in the
center of the lead) is also shown.  The coordinates of the four
corners of the right lead, as well as the coordinates of the points at
which the two wires are connected to the right lead, are indicated. As
shown, we always assume that the wires are attached near the center of
the short edges of the leads. }
\label{coordinates}
\end{figure*}

In the light of the foregoing remarks, the multiple-connectedness of
the device suggests that one should anticipate oscillations with
magnetic field, e.g., in the device resistance. Oscillations are
indeed observed. But they are distinct from the resistance
oscillations observed by Little and Parks and from the critical
current oscillations observed in SQUID rings.  What distinguishes the
resistance oscillations reported in Ref.~\cite{us_in_science} from
those found, e.g., by Little and Parks?  First, the most notable
aspect of these oscillations is the value of their period.  In the
Little-Parks type of experiment, the period is given by \(\Phi_0/ 2 a
b\), where \(\Phi_0(\equiv h c / 2 e)\) is the superconducting flux
quantum, \(2 a\) is the bridge separation, and \(b\) is the bridge
length, i.e., the superconducting flux quantum divided by the area of
the AB contour (see Fig.~\ref{coordinates}). In a high-magnetic-field
regime, such periodic behavior is indeed observed experimentally, with
the length of the period somewhat shorter but of the same order of
magnitude as in the AB effect~\cite{us_in_science}. However, in a
low-magnetic-field regime, the observed period is appreciably smaller
(in fact by almost two orders of magnitude for our device geometry).
Second, because the resistance is caused by thermal phase fluctuations
(i.e.~phase slips) in very narrow wires, the oscillations are
observable over a wide range of temperatures (\(\sim
1\,\text{K}\)). Third, the Little-Parks resistance is wholly ascribed
to a rigid shift of the \(R(T)\) curve with magnetic field, as
\(T_\text{c}\) oscillates.  In contrast, in our system we observe a
periodic broadening of the transition (instead of the
Little-Parks---type rigid shift) with magnetic field. Our theory
explains quantitatively this broadening via the modulation of the
barrier heights for phase slips of the superconducting order parameter
in the nanowires.

In the experiment, the sample is cooled in zero magnetic field, and
the field is then slowly increased while the resistance is
measured. At a sample-dependent field (\(\sim 5\,\text{mT}\)) the
behavior switches sharply from a low-field to a high-field regime. If
the high-field regime is not reached before the magnetic field is
swept back, the low-field resistance curve is reproduced. However,
once the high-field regime has been reached, the sweeping back of the
field reveals phase shifts and hysteresis in the \(R(B)\) curve.  The
experiments~\cite{us_in_science} mainly address rectangular leads that
have one mesoscopic and one macroscopic dimension. Therefore, we shall
concentrate on such strip geometries. We shall, however, also discuss
how to extend our approach to generic (mesoscopic) lead shapes.  We
note in passing that efficient numerical methods, such as the boundary
element method (BEM)~\cite{BEM}, are available for solving the
corresponding Laplace problems.

This paper is arranged as follows. In Section~\ref{sec_phys} we
construct a basic picture for the period of the magnetoresistance
oscillations of the two-wire device, which shows how the mesoscopic
size of the leads accounts for the anomalously short magnetoresistance
period in the low-field regime.  In Section \ref{sec:leads} we
concentrate on the properties of mesoscopic leads with regard to their
response to an applied magnetic field, and in Section
\ref{sec:bridges} we extend the LAMH model to take into account the
inter-wire coupling through the leads. Analytical expressions are
derived for the short- and long-wire limits, whilst a numerical
procedure is described for the general case. The predictions of the
model are compared with data from our experiment in Section
\ref{sec:experiment}, and we give some concluding remarks in Section
\ref{sec:conclusion}. Certain technical components are relegated to
the appendix, as is the analysis of example multiwire devices.

\section{Origin of magnetoresistance oscillations}
\label{sec_phys}
Before presenting a detailed development of the theory, we give an
intuitive argument to account for the anomalously-short period of the
magnetoresistance in the low-magnetic-field regime, mentioned above.  

\subsection{Device geometry}
The geometry of the devices studied experimentally is shown in
Fig.~\ref{coordinates}.  Five devices were successfully fabricated and
measured.  The dimensions of these devices are listed in
Table~\ref{table:period}, along with the short magnetoresistance
oscillation period.  The perpendicular penetration depth
\(\lambda_\perp\) for the films used to make the leads is roughly
\(70\,\mu\text{m}\), and coherence length \(\xi\) is roughly \(5 \,
\text{nm}\).

\subsection{Parametric control of the state of the wires by the leads}
\begin{figure}
\includegraphics[width=8cm]{\figdir/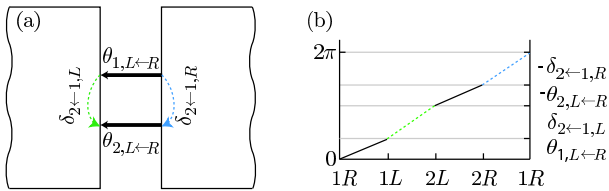}
\caption{\label{loop} (a)~Close-up of the two nanowires and the
leads. The top (bottom) thick arrow represents the integration contour
for determining the phase accumulation \(\theta_{1, L \leftarrow R}\)
(\(\theta_{2, L \leftarrow R}\)) in the first (second) wire.  The
dotted arrow in the left (right) lead indicates a possible choice of
integration contour for determining the phase accumulation \(\delta_{2
\leftarrow 1, L}\) (\(\delta_{2 \leftarrow 1, R}\)).  These contours
may be deformed without affecting the values of the various phase
accumulations, as long as no vortices are crossed. (b)~Sketch of the
corresponding superconducting phase at different points along the AB
contour when one vortex is located inside the contour.}
\end{figure}
The essential ingredients in our model are (i)~leads, in which the
applied magnetic field induces supercurrents and hence gradients in
the phase of the order parameter, and (ii)~the two wires, whose
behavior is controlled parametrically by the leads through the
boundary conditions imposed by the leads on the phase of the order
parameters in the wires.  For now, we assume that the wires have
sufficiently small cross-sections that the currents through them do
not feed back on the order parameter in the leads.  (In
Section~\ref{sec:blcouple} we shall discuss when this assumption may
be relaxed without altering the oscillation period.)\thinspace\ The
dissipation results from thermally activated phase slips, which cause
the superconducting order parameter to explore a discrete family of
local minima of the free energy.  (We assume that the barriers
separating these minima are sufficiently high to make them
well-defined states.)\thinspace\ These minima (and the saddle-point
configurations connecting them) may be indexed by the net
(i.e.~forward minus reverse) number of phase slips that have occurred
in each wire (\(n_1\) and \(n_2\), relative to some reference
state). More usefully, they can be indexed by \(n_s = \min(n_1, n_2)\)
(i.e.~the net number of phase slips that have occurred in both wires)
and \(n_v = n_1-n_2\) (i.e.~the number of vortices enclosed by the AB
contour, which is formed by the wires and the edges of the leads).  We
note that two configurations with identical \(n_v\) but distinct
\(n_s\) and \(n_s'\) have identical order parameters, but differ in
energy by
\begin{equation}
\int I V \, dt=\frac{\hbar}{2 e} \int I \dot{\Theta} \, dt =
\frac{h}{2 e} \, I \, (n_s' - n_s),
\end{equation}
due to the work done by the current source supplying the current
\(I\), in which \(V\) is the inter-lead voltage, \(\Theta\) is the
inter-lead phase difference as measured between the two points
half-way between the wires, and the Josephson relation
\(\dot{\Theta}=2 e V/ \hbar\) has been invoked.  In our model, we
assume that the leads are completely rigid. Therefore the rate of
phase change, and thus the voltage, is identical at all points inside
one lead. For sufficiently short wires, \(n_v\) has a unique value, as
there are no stable states with any other number of vortices.

Due to the screening currents in the left lead, induced by the applied
magnetic field \(B\) (and independent of the wires), there is a
field-dependent phase \(\delta_{2 \leftarrow 1, L}(B)=\int_1^2
d\vec{r}\cdot \vec{\nabla}\varphi(B)\) (computed below) accumulated in
passing from the point at which wire 1 (the top wire) contacts the
left (L) lead to the point at which wire 2 (the bottom wire) contacts
the left lead (see Fig.~\ref{loop}). Similarly, the field creates a
phase accumulation \(\delta_{2 \leftarrow 1, R}(B)\) between the
contact points in the right (R) lead.  As the leads are taken to be
geometrically identical, the phase accumulations in them differ in
sign only: \(\delta_{2 \leftarrow 1, L}(B) = -\delta_{2 \leftarrow 1,
R}(B)\).  We introduce \(\delta(B) = \delta_{2 \leftarrow 1, L}(B)\).
In determining the local free-energy minima of the wires, we solve the
Ginzburg-Landau equation for the wires for each vortex number \(n_v\),
imposing the single-valuedness condition on the order parameter,
\begin{equation}
\theta_{1, L \leftarrow R}-\theta_{2, L \leftarrow R}+2 \delta(B)=2
\pi n_v.
\label{eq:phaseconstraint}
\end{equation}
This condition will be referred to as the {\it phase constraint}\/.
Here, \(\theta_{1, L \leftarrow R}=\int_R^L d\vec{r}\cdot
\vec{\nabla}\varphi(B)\) is the phase accumulated along wire 1 in
passing from the right to the left lead; \(\theta_{2, L \leftarrow
R}\) is similarly defined for wire 2.

Absent any constraints, the lowest energy configuration of the
nanowires is the one with no current through the wires. Here, we adopt
the gauge in which \(\boldsymbol{A}= B y \boldsymbol{e}_x\) for the
electromagnetic vector potential, where the coordinates are as shown
in Fig.~\ref{coordinates}.  The Ginzburg-Landau expression for the
current density in a superconductor is
\begin{equation}
\boldsymbol{J} \propto \left(\boldsymbol{\nabla} \varphi(\boldsymbol{r}) -
\frac{2 e}{\hbar} \boldsymbol{A}(\boldsymbol{r}) \right).
\label{currentD0}
\end{equation}
For our choice of gauge, the vector potential is always parallel to
the nanowires, and therefore the lowest energy state of the nanowires
corresponds to a phase accumulation given by the flux through the AB
contour, \(\theta_{1, L \leftarrow R}=-\theta_{2, L \leftarrow R}=2
\pi B a b/\Phi_0\).  As we shall show shortly, for our device geometry
(i.e.~when the wires are sufficiently short, i.e., \(b \ll l\)), this
phase accumulation may be safely ignored, compared to the phase
accumulation \(\delta(B)\) associated with screening currents induced
in the leads.  As the nanowires are assumed to be weak compared to the
leads, to satisfy the phase constraint~(\ref{eq:phaseconstraint}), the
phase accumulations in the nanowires will typically deviate from their
optimal value, generating a circulating current around the AB contour.
As a consequence of LAMH theory, this circulating current results in a
decrease of the barrier heights for phase slips, and hence an increase
in resistance.  The period of the observed oscillations is derived
from the fact that whenever the magnetic field satisfies the relation
\begin{equation}
2 \pi m = 
2 \pi \frac{2 a b B}{\Phi_0} + 2 \delta(B)
\label{eq:period}
\end{equation}
[where \(m\) is an integer and the factor of \(2\) accompanying
\(\delta(B)\) reflects the presence of two leads], there is no
circulating current in the lowest in energy state, resulting in
minimal resistance.  Furthermore, the family of free energy-minima of
the two-wire system (all of which, in thermal equilibrium, are
statistically populated according to their energies) is identical to
the \(B=0\) case. The mapping between configurations at zero and
nonzero \(B\) fields is established by a shift of the index \(n_v
\rightarrow n_v - m\). Therefore, as the sets of physical states of
the wires are identical whenever the periodicity
condition~(\ref{eq:period}) is satisfied, at such values of \(B\) the
resistance returns to its \(B=0\) value.

\subsection{Simple estimate of the oscillation period}
\begin{figure}
\fbox{\includegraphics[width=6cm]{\figdir/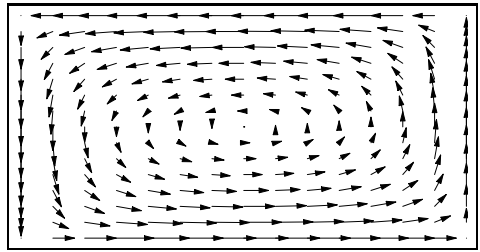}}
\caption{Current profile in a long superconducting strip, calculated
for a finite-length strip by summing the series for
\(\boldsymbol{\nabla} \phi- \frac{2 \pi}{\Phi_0} \boldsymbol{A}\)
[from \Eqn{eq_phi}] numerically. Note that there is no vortex in the
center of the lead.}
\label{2dj}
\end{figure}

In this subsection, we will give a ``back of the envelope'' estimate
for the phase gain \(\delta(B)\) in a lead by considering the current
and phase profiles in one such lead.  According to the Ginzburg-Landau
theory, in a mesoscopic superconductor, subjected to a weak magnetic
field, the current density is given by \Eqn{currentD0}.  Now consider
an isolated strip-shaped lead used in the device. Far from either of
the short edges of this lead, \(\boldsymbol{A}= B y \boldsymbol{e}_x\)
is a London gauge~\cite{London}, i.e., along all surfaces of the
superconductor \(\boldsymbol{A}\) is parallel to them;
\(\boldsymbol{A}\rightarrow 0\) in the center of the superconductor;
and \(\boldsymbol{\nabla}\cdot \boldsymbol{A}=0\).  In this special
case, the London relation~\footnote{Consider the case in which
\(\boldsymbol{A}\) is a London gauge everywhere (with our choice of
gauge, \(\boldsymbol{A}= B y \boldsymbol{e}_x\), this is the case for
an infinitely long strip). By using the requirement that
\(\boldsymbol{\nabla}\cdot \boldsymbol{A}=0\), together with
\Eqn{GLimg}, we see that \(\phi\) satisfies the Laplace equation.  We
further insist that no current flows out of the superconductor, i.e.,
along all surfaces the supercurrent density, \Eqn{currentD}, is always
parallel to the surface. Together with the requirement that along all
surfaces \(\boldsymbol{A}\) is parallel to them, this implies the
boundary condition that \(\boldsymbol{n} \cdot \boldsymbol{\nabla}
\phi=0\). Next, it can be shown that this boundary condition implies
that \(\phi\) must be a constant function of position in order to
satisfy the Laplace equation, and therefore~\Eqn{currentD} simplifies
to read \(\boldsymbol{J} = -(c/8 \pi \lambda_\text{eff}^2)
\boldsymbol{A}\), which is known as the London relation.}
states that the supercurrent density is proportional to the vector
potential in the London gauge.  Using this relation, we find that the
supercurrent density is \(\boldsymbol{J} \propto - (2e / \hbar) \,
\boldsymbol{A} = -(2e/\hbar) \, B y \boldsymbol{e}_x\), i.e., there is
a supercurrent density of magnitude \(\propto (2e/\hbar) \, B l\)
flowing to the left at the top (long) edge of the strip and to the
right at the bottom (long) edge.  At the two short ends of the strip,
the two supercurrents must be connected, so there is a supercurrent
density of magnitude \(\sim (2e / \hbar) B l\) flowing down the left
(short) edge of the strip and up the right (short) edge (see
Fig.~\ref{2dj}).
Near the short ends of the strips, our choice of gauge no longer
satisfies the criteria for being a London gauge, and therefore
\(\boldsymbol{\nabla} \phi\) may be nonzero. As, in our choice of
gauge, \(\boldsymbol{A}\) points in the \(\boldsymbol{e}_x\)
direction, the supercurrent on the ends of the strip along
\(\boldsymbol{e}_y\) must come from the \(\nabla_y \phi\) term. Near
the center of the short edge \(\nabla_y \phi = - 2 \pi c_1 l/\Phi_0
B\). The phase difference between the points \((-L,-a)\) and
\((-L,a)\) is therefore given by
\begin{equation}
\delta(B)=\int_{-a}^a \nabla_y \phi \, dy = - \frac{2 \pi c_1}{\Phi_0}
B \, 2 a l,
\label{eq:approx_delta_B}
\end{equation}
where we have substituted \(2\pi/\Phi_0\) for \(2e/\hbar\) and
\(c_1(a/l)\) is a function of order unity, which accounts for how the
current flows around the corners.  As we shall show, \(c_1\) depends
only weakly on \(a/l\), and is constant in the limit \(a \ll l\).

Finally, we obtain the magnetoresistance period by substituting
\Eqn{eq:approx_delta_B} into \Eqn{eq:period}:
\begin{align}
\Delta B = \left[ \left(\frac{\Phi_0}{c_1\, 4 a l}\right)^{-1}
+\left(\frac{\Phi_0}{2 a b}\right)^{-1} \right]^{-1} \, .
\label{p1}
\end{align}
Thus, we see that for certain geometries the period is largely
determined not by the flux threading through the geometric area \(2 a
b\) but by the response of the leads and the corresponding effective
area \(4 a l\), provided the nanowires are sufficiently short
(i.e.~\(b \ll l\)), justifying our assumption of ignoring the phase
gradient induced in the nanowires by the magnetic field.

In fact, we can also make a prediction for the periodicity of the
magnetoresistance at high magnetic fields, i.e., when vortices have
penetrated the leads (see Section~\ref{sec:likharev}). To do this, we
should replace \(l\) in~\Eqn{p1} by the characteristic inter-vortex
spacing \(r\).  Note that if \(r\) is comparable to \(b\), we can no
longer ignore the flux through the AB contour.  Furthermore, if \(r
\ll b\) then the flux through the AB contour determines periodicity
and one recovers the usual Aharonov-Bohm type of phenomenology.

\section{Mesoscale superconducting leads}
\label{sec:leads}

In this section and the following one we shall develop a detailed
model of the leads and nanowires that constitute the mesoscopic
device.

\subsection{Vortex-free and vorticial regimes}
\label{sec:likharev}

Two distinct regimes of magnetic field are expected, depending on
whether or not there are trapped (i.e.~locally stable) vortices inside
the leads.  As described by Likharev~\cite{Likharev}, a vortex inside
a superconducting strip-shaped lead is subject to two forces.
First, due to the the currents induced by the magnetic field there is
a Magnus force pushing it towards the middle of the strip.  Second,
there is a force due to image vortices (which are required to enforce
the boundary condition that no current flows out of the strip and into
the vacuum) pulling the vortex towards the edge.  When the two forces
balance at the edge of the strip, there is no energy barrier
preventing vortex penetration and vortices enter.  Likharev has
estimated of the corresponding critical magnetic field to be
\begin{equation}
H_{\rm s} \approx \frac{\Phi_0}{\pi d} \frac{1}{\xi a(1)},
\end{equation}
where \(d(\equiv2 l)\) is the width of the strip and \(a(1) \sim 1\)
for strips that are much narrower than the penetration depth (i.e.~for
\(d \ll \lambda\)).

Likharev has also shown that, once inside a strip, vortices remain
stable inside it down to a much lower magnetic field \(H_{c1}\),
given by
\begin{equation}
H_\text{c1}= \frac{\Phi_0}{\pi d} \frac{2}{d} \ln \left(
\frac{d}{4\xi} \right).
\end{equation}
At fields above \(H_\text{c1}\) the potential energy of a vortex
inside the strip is lower than for one outside (i.e.~for a virtual
vortex~\cite{ref:topology_note}).  Therefore, for magnetic fields in
the range \(H_\text{c1}<H<H_\text{s}\) vortices would remain trapped
inside the strip, but only if at some previous time the field were
larger than \(H_\text{s}\).  This indicates that hysteresis with respect to
magnetic field variations should be observed, once \(H\) exceeds
\(H_\text{s}\) and vortices become trapped in the leads.

In real samples, in addition to the effects analyzed by Likharev,
there are also likely to be locations (e.g.~structural defects) that
can pin vortices, even for fields smaller than \(H_{c1}\), so the
reproducibility of the resistance {\it vs.\/}~field curve is not
generally expected once \(H_\text{s}\) has been surpassed.

As magnetic field at which vortices first enter the leads is
sensitive to the properties of their edges, we expect only rough
agreement with Likharev's theory.  For sample 219-4, using Likharev's
formula, we estimate \(H_\text{s}=11\,\text{mT}\) (with
\(\xi=5\,\text{nm}\)). The change in regime from fast to slow
oscillations is found to occur at \(3.1\,\text{mT}\) for that
sample~\cite{us_in_science}.  It is possible to determine the critical
magnetic fields \(H_\text{s}\) and \(H_\text{c1}\) by the direct
imaging of vortices. Although we do not know of such a direct
measurement of \(H_\text{s}\), \(H_\text{c1}\) was determined by field
cooling niobium strips, and found to agree in magnitude to Likharev's
estimate~\cite{Martinis2004}.

\subsection{Phase variation along the edge of the lead}
\label{sec_exact}

In the previous section it was shown that the periodicity of the
magnetoresistance is due to the phase accumulations associated with
the currents along the edges of the leads between the nanowires.
Thus, we should make a precise calculation of the dependence of these
currents on the magnetic field, and this we now do.

\subsubsection{Ginzburg-Landau theory}
To compute \(\delta(B)\), we start with the Ginzburg-Landau equation
for a thin film as our description of the mesoscopic superconducting
leads:
\begin{equation}
\alpha \psi+ \beta |\psi|^2 \psi +
\frac{1}{2 m^*} \left(\frac{\hbar}{i} \boldsymbol{\nabla} -
\frac{e^*}{c} \boldsymbol{A}\right)^2 \psi = 0.
\label{GLfree}
\end{equation}
Here, \(\psi\) is the Ginzburg-Landau order parameter, \(e^*\)
(\(=2e\)) is the charge of a Cooper pair and \(m^*\) is its mass, and
\(\alpha\) and \(\beta\) may be expressed in terms of the coherence
length \(\xi\) and critical field \(H_\text{c}\) via
\(\alpha=-\hbar^2/2 m^* \xi^2\) and \(\beta=4 \pi
\alpha^2/H_\text{c}^2\).

The assumptions that the magnetic field is sufficiently weak and that
the lead is a narrow strip (compared with the magnetic penetration
depth) allow us to take the {\it amplitude\/} of the order parameter
in the leads to have the value appropriate to an infinite thin film in
the absence of the field.  By expressing the order parameter in terms
of the (constant) amplitude \(\psi_0\) and the (position-dependent)
phase \(\phi(\boldsymbol{r})\), i.e.,
\begin{equation}
\psi(\boldsymbol{r})=\psi_0 \, e^{i \phi(\boldsymbol{r})}, 
\end{equation}
the Ginzburg-Landau formula for the current density, 
\begin{equation}
\boldsymbol{J}=
\frac{e^* \hbar}{2 m^* i}
\big(\psi^{\ast} \boldsymbol{\nabla} \psi
-\psi\boldsymbol{\nabla} \psi^{\ast}\big)
-\frac{{e^*}^2}{m^* c}\psi^{\ast}\psi 
\boldsymbol{A}(\boldsymbol{r}),
\label{current-old}
\end{equation}
becomes 
\begin{equation}
\boldsymbol{J}=\frac{e^*}{m^*}\psi_0^2 \big( \hbar \boldsymbol{\nabla}
\phi(\boldsymbol{r}) - \frac{e^*}{c} \boldsymbol{A}(\boldsymbol{r})
\big),
\label{currentD}
\end{equation}
and [after dividing by \(e^{i \phi(\boldsymbol{r})}\)] the real and
imaginary parts of the Ginzburg-Landau equation become
\begin{subequations}
\begin{align}
0&=\left[\alpha \, \psi_0 + \beta \, \psi_0^3 + \frac{1}{2 m^*} \psi_0
\left| \hbar \boldsymbol{\nabla} \phi(\boldsymbol{r}) - \frac{e^*}{c}
\boldsymbol{A}(\boldsymbol{r})\right|^2\right], \label{eq13a}\\
0&=\frac{\hbar^2}{2 m^* i} \, \psi_0 \left(\nabla^2 \phi(\boldsymbol{r}) -
\frac{e^*}{\hbar c} \boldsymbol{\nabla} \cdot
\boldsymbol{A}(\boldsymbol{r}) \right).
\label{GLimg}
\end{align}
\end{subequations}
As long as any spatial inhomogeneity in the gauge-covariant derivative
of the phase is weak on the length-scale of the coherence length
\(\big[\)~i.e.~\(\xi \big|\boldsymbol{\nabla} \phi(\boldsymbol{r}) -
\frac{e^*}{\hbar c} \boldsymbol{A}(\boldsymbol{r})\big| \ll
1\)~\(\big]\), the third term in \Eqn{eq13a} is much smaller than the
first two and may be ignored, fixing the amplitude of the order
parameter at its field-free infinite thin film value, {\it viz.\/},
\(\bar{\psi_0}\equiv\sqrt{-\alpha/\beta}\). To compute
\(\phi(\boldsymbol{r})\) we need to solve the imaginary part of the
Ginzburg-Landau equation.

\subsubsection{Formulation as a Laplace problem}
We continue to work in the approximation that the amplitude of the
order parameter is fixed at \(\bar{\psi_0}\). Starting from
\Eqn{GLimg}, we see that for our choice of gauge, \(\boldsymbol{A}= B
y \boldsymbol{e}_x\), the phase of the order parameter satisfies the
Laplace equation, \(\nabla^2 \phi=0\). We also enforce the boundary
condition that no current flows out of the superconductor on boundary
surface $\Sigma$, whose normal is ${\bf n}$:
\begin{subequations}
\label{eq:bc}
\begin{align}
&\boldsymbol{n}\cdot \boldsymbol{j} \big|_{\Sigma}=0,\\
&\boldsymbol{j}\propto\Big(\boldsymbol{\nabla} \phi - \frac{2
\pi}{\Phi_0} \boldsymbol{A} \Big).
\end{align} 
\end{subequations}

\subsubsection{Solving the Laplace problem for the strip geometry}

To solidify the intuition gained via the physical arguments given in
Section~\ref{sec_phys}, we now determine the phase profile for an
isolated superconducting strip in a magnetic field. This will allow us
to determine the constant \(c_1\) in \Eqn{p1}, and hence obtain a
precise formula for the magnetoresistance period.  To this end, we
solve Laplace's equation for \(\phi\) subject to the boundary
conditions~(\ref{eq:bc}).  We specialize to the case of a rectangular
strip
\footnote{This specialization is not necessary, but it is convenient and
adequately illustrative}.

In terms of the coordinates defined in Fig.~\ref{coordinates}, we expand
\(\phi(x,y)\) as the superposition
\begin{equation}
\phi(x,y)=\Theta_{\text{L/R}}+\sum_k 
\left(A_k \, e^{-k x} + B_k \, e^{k x}\right)
\sin(k y), 
\label{eq_phi}
\end{equation}
which automatically satisfies Laplace's equation, although the boundary
conditions remain to be satisfied. \(\Theta_{\text{L(R)}}\) is the
phase at the the point in the left (right) lead located half-way
between the wires. In other words \(\Theta_L=\phi(-L-b,0)\) and
\(\Theta_R=\phi(-L,0)\) in the coordinate system indicated in
Fig.~\ref{coordinates}.  \(\Theta_{\text{L/R}}\) are not determined by
the Laplace equation and boundary conditions, but will be determined
later by the state of the nanowires.

\begin{figure}
\includegraphics[width=8cm]{\figdir/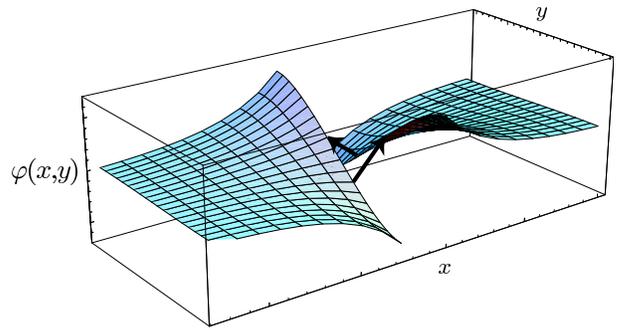}
\caption{Phase profile in the leads in the vicinity of the trench,
generated by numerically summing the series for \(\phi\) for a
finite-length strip. Arrows indicate phases connected by nanowires. }
\label{3dplots}
\end{figure}

We continue working in the gauge \(\boldsymbol{A}=B y \,
\boldsymbol{e}_x\).  The boundary conditions across the edges at
\(y=\pm l\) (i.e.~the long edges) are \(\partial_y \phi(x,y=\pm
l)=0\).  These conditions are satisfied by enforcing \(k_n=\pi
(n+\frac{1}{2})/l\), where \(n=0,1,2,\ldots\).  The boundary
conditions across the edges at \(x=\pm L\) (i.e.~the short edges) are
\(\partial_x \phi(x=\pm L,y)=h y\) (where \(h \equiv 2 \pi
B/\Phi_0\)).  This leads to the coefficients in \Eqn{eq_phi} taking
the values
\begin{align}
B_k&=-A_k=\frac{h}{k_n^3 l} \frac{(-1)^n}{\cosh(k_n L)} & (n=0,1,\ldots),
\end{align}
and hence to the solution
\begin{align}
\phi(x,y)=\sum_{n=0}^\infty 
\frac{(-1)^n\,2 h}{k_n^3 l \cosh(k_nL)}\,
\sin(k_n y)\,\sinh(k_n x).
\end{align}
Figure~\ref{3dplots} shows the phase profiles in the leads, in the
region close to the trench that separates the leads.

\subsection{Period of magnetoresistance for leads having a 
rectangular strip geometry}
\label{subsec:period} 
\begin{figure}
\includegraphics{\figdir/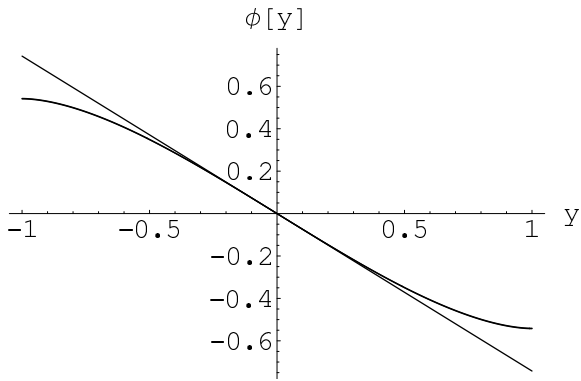}
\caption{Phase profile on the \(x=-L\) (i.e.~short) edge of the strip.
Numerical summation [Eq.~(\ref{eq:nonlinear}) with 100 terms] for
\(L=2\) and \(L=\infty\) (\(l=1)\) as well as the linear form from
Eq.~(\ref{lin_fit}).  Note that the linear fit is good near the origin
(e.g.~for \(a \alt 0.25 l\)), and the curves for \(L=2\) and
\(L=\infty\) coincide.}
\label{fit_plot}
\end{figure}

Using the result for the phase that we have just established, we see
that the phase profile on the short edge of the strip at \(x=-L\) is
given by
\begin{equation}
\label{eq:nonlinear}
\phi(-L,y)= - \frac{2 h l^2}{\pi^2}\sum_{n=0}^\infty
\frac{(-1)^n}{(n+\frac{1}{2})^3} \sin 
\frac{\pi\left(n+\frac{1}{2}\right) y}{l},
\end{equation}
where we have taken the limit \(L\rightarrow \infty\).  We would like
to evaluate this sum at the points \((x,y)=(-L,\pm a)\). This can be
done numerically. For nanowires that are close to each other (i.e.~for
\(a \ll l\)), an approximate value can be found analytically by
expanding in a power series in $a$ around \(y=0\):
\begin{equation}
\begin{split}
\phi(-L,a)=\phi(-L,0)+a \, \left. \frac{\partial}{\partial y}
\phi(-L,y) \right|_{y=0} \quad\quad
\\
+\frac{a^2}{2}
\left. \frac{\partial^2}{\partial y^2} \phi(-L,y) \right|_{y=0} +
O(a^3).
\end{split}
\end{equation}
The first and third terms are evidently zero, as \(\phi\) is an odd
function of \(y\). The second term can be evaluated by changing the
order of summation and differentiation.  (Higher-order
terms are harder to evaluate, as the changing of the order of
summation and differentiation does not work for them.)\thinspace\  
Thus, to leading order in $a$ we have 
\begin{align}
\phi(-L,a)\approx-\frac{8 G}{\pi^2} h l a, 
\label{lin_fit}
\end{align}
where \(G\equiv\sum_{n=0}^\infty \frac{(-1)^n}{(2n+1)^2}\approx
0.916\) is the Catalan number (see Ref.~\cite{catalan_n}).  This
linear approximation is plotted, together with the actual phase profile
obtained by the numerical evaluation of Eq.~(\ref{eq:nonlinear}), in
Fig.~\ref{fit_plot}.  Hence, the value of \(c_1\) in
\Eqn{eq:approx_delta_B} becomes \(c_1=8 G /\pi^2 \approx 0.74\),
and \Eqn{p1} becomes
\begin{align}
B=\frac{\Phi_0}{2 \pi}h= 
\frac{\pi^2}{8 G} \frac{\Phi_0}{4 a l}.
\end{align}
To obtain this result we used the relation \(\delta(B)/2=\phi(-L,a)\).

\subsection{Bridge-lead coupling}
\label{sec:blcouple}
In order to simplify our analysis we have assumed that the nanowires
do not exert any influence on the order parameter in the leads.  We
examine the justification for this assumption in the setting of the
experiment that we are attempting to describe~\cite{us_in_science}.

The assumption will be valid if the bending of the phase of the order
parameter, in order to accommodate any circulating current around
the AB contour, occurs largely in the nanowires.
As the phase of the order parameter in the leads satisfies the Laplace
equation, which is linear, we can superpose the circulating-current
solution with the previously-obtained magnetic-field-induced
solution. The boundary conditions on the right lead for the
circulating-current solution are \(\boldsymbol{n} \cdot
\boldsymbol{\nabla} \phi=0\) everywhere, except at the two points
where the nanowires are attached to the lead [i.e.~at \((x,y)=(L,\pm
a)\)]. Treating the nanowires as point current sources, the boundary
condition on the short edge of the right lead is
\(\partial_x \phi = I (\Phi_0/H_\text{c}^2 s \xi)
(\delta(y+a)-\delta(y-a)\)), where \(I\) is the current circulating in
the loop, and \(H_\text{c}\), \(s\), \(\xi\) are the film critical field,
thickness, and coherence length. By using the same expansion as before,
\Eqn{eq_phi}, we obtain the coefficients of the
Fourier-series in the long strip limit:
\begin{align}
A_k=I (\Phi_0/H_\text{c}^2 s \xi) \frac{2 \sin(ka)}{k l \exp(k l)}.
\end{align}
Having the coefficients of the Fourier series, we can find the phase
difference in the right lead between the two points at which the
nanowires connect to the right lead, induced in this lead by the
current circulating in the loop:
\begin{align}
\delta_{cc}=2 I (\Phi_0/H_\text{c}^2 s \xi) \sum_{n=0}^{k=1/w} \frac{\sin^2(k
a)}{k l} \sim \ln(2l/\pi w).
\end{align}
Here, we have introduced a large wave-vector \(k\) cut-off at the
inverse of the width \(w\) of the wire.  On the other hand, the
current flowing through the wire is
\begin{equation}
\frac{\xi H_\text{c}^2}{\Phi_0} w s \frac{\Delta \theta}{b},
\end{equation}
where \(\xi\), \(H_\text{c}\), and \(s\) are the wire coherence length,
critical field, and height (recall that \(b\) is the wire length). To
support a circulating current that corresponds to a phase accumulation
of \(\Delta \theta\) along one of the wires, the phase difference
between the two nanowires in the lead must be on the order of
\begin{equation}
\delta_{cc}=\Delta \theta \frac{w}{b} 
\frac{\left( H_\text{c}^2 s \xi \right)_\text{wire}}
{\left( H_\text{c}^2 s \xi \right)_\text{film}}
\ln\left(\frac{2 l}{\pi w}\right).
\end{equation}
For our experiments~\cite{us_in_science}, we estimate that the ratio of
\(\delta_{cc}\) to \(\Delta \theta\) is always less than \(20\%\),
validating the assumption of weak coupling.

\subsection{Strong nanowires} 
We remark that the assumption of weak nanowires {\it is not
obligatory\/} for the computation the magnetoresistance {\it
period\/}.  Dropping this assumption would leave the period of the
magnetoresistance oscillations unchanged.

To see this, consider \(\phi_{11}\), i.e., the phase profile in the
leads that corresponds to the lowest energy solution of the
Ginzburg-Landau equation at field corresponding to the first
resistance minimum [i.e.~at B being the first non-zero solution of
\Eqn{eq:period}].  For this case, and for short wires, the phase gain
along the wires is negligible, whereas the phase gain in the leads is
\(2 \pi\), even for wires with large critical current.  Excited
states, with vortices threading the AB contour, can be constructed by
the linear superposition of \(\phi_{11}\) with \(\phi_{0n_v}\), where
\(\phi_{0n_v}\) is the phase profile with \(n_v\) vortices at no
applied magnetic field.

This construction requires that the nanowires are narrow, but works
independently of whether nanowires are strong or weak, in the limit
that \(H \ll H_\text{c}\).  The energy of the lowest energy state always
reaches its minimum when the applied magnetic field is such that there
is no phase gain (i.e.~no current) in the nanowires.  By the above
construction, it is clear that the resistance of the device at this
field is the same as at zero field, and therefore the minimum
possible.

Therefore, our calculation of the period is valid, independent of
whether the nanowires are weak or strong.  However, the assumption of
weak nanowires is necessary for the computation of magnetoresistance
amplitude, which we present in the following section.

\section{Parallel superconducting nanowires and intrinsic resistance} 
\label{sec:bridges}

In this section we consider the intrinsic resistance of the device. We
assume that this resistance is due to thermally activated phase slips
(TAPS) of the order parameter, and that these occur within the
nanowires. Equivalently, these processes may be thought of as
thermally activated vortex flow across the nanowires.  Specifically,
we shall derive analytical results for the asymptotic cases of
nanowires that are either short or long, compared to coherence length,
i.e.~Josephson junctions~\cite{IZ, AH} or
Langer-Ambegaokar-McCumber-Halperin (LAMH) wires~\cite{LA,MH}; see
also Ref.~\cite{little}.  We have not been able to find a closed-form
expression for the intrinsic resistance in the intermediate-length
regime, so we shall consider that case numerically.

There are two (limiting) kinds of experiments that may be performed:
fixed total current and fixed voltage. In the first kind, a specified
current is driven through the device and the time-averaged voltage is
measured. Here, this voltage is proportional to the net number of
phase slips (in the forward direction) per unit time, which depends on
the height of the free-energy barriers for phase slips.  Why do we
expect minima in the resistance at magnetic fields corresponding to
\(2\delta=2 m \pi\) and maxima at \(2\delta=(2m+1)\pi\) for \(m\)
integral, at least at vanishingly small total current through both
wires?  For \(2\delta=2 m \pi\) the nanowires are unfrustrated, in the
sense that there is no current through either wire in the lowest local
minimum of the free energy.  On the other hand, for
\(2\delta=(2m+1)\pi\) the nanowires are maximally frustrated: there is
a nonzero circulating current around the AB contour. Quite generally,
the heights of the free-energy barriers protecting locally stable
states decrease with increasing current through a wire, and thus the
frustrated situation is more susceptible to dissipative fluctuations,
and hence shows higher resistance. Note, however, that due to the
inter-bridge coupling caused by the phase constraint, the resistance
of the full device is more subtle than the mere addition of the
resistances of two independent, parallel nanowires, both carrying the
requisite circulating current.

In the second kind of experiment, a fixed voltage is applied across
the device and the total current is measured. In this situation, the
inter-lead voltage is fixed, and therefore the phase drop along each
wire is a fixed function of time. Hence, there is no inter-bridge
coupling in the fixed voltage regime. Therefore, the resistance of the
device would not exhibit magnetic field dependence. If the voltage is
fixed far away from the wires, but not in the immediate vicinity of
the wires, so that the phase drop along each wire is not rigidly
fixed, then some of the magnetic field dependence of the resistance
would be restored.  In our experiments on two-wire devices, we believe
that the situation lies closer to the fixed current limit than to the
fixed voltage limit, and therefore we shall restrict our attention to
the former limit.

In the fixed-current regime, the relevant independent thermodynamic
variable for the device is the total current through the pair of
wires, i.e., \(I\equiv I_1+I_2\). Therefore, the appropriate free
energy to use, in obtaining the barrier heights for phase slips, is
the Gibbs free energy \(G(I)\), as discussed by McCumber~\cite{M68},
rather than the Helmholtz free energy \(F(\Theta)\)~\footnote{Recall
that the Helmholtz free energy is obtained by minimizing the
Ginzburg-Landau free energy functional with respect to the order
parameter function \(\psi(\boldsymbol{r})\), subject to the phase
accumulation constraint \(\int_\text{L}^\text{R} d\boldsymbol{r} \cdot
\boldsymbol{\nabla} \phi = \theta\).}. In the Helmholtz free energy
the independent variable can be taken to be \(\Theta \equiv
\Theta_\text{L}-\Theta_\text{R}\), i.e., the phase difference across
the center of the ``trench,'' defined modulo \(2 \pi\). \(G(I)\) is
obtained from \(F(\Theta)\) via the appropriate Legendre
transformation:
\begin{equation}
G(I)=F(\Theta)-\frac{\hbar}{2 e} I \Theta,
\end{equation}
where the second term represents the work done on the system by the
external current source. \(F(\Theta)\) is the sum of the Helmholtz
free energies for the individual nanowires:
\begin{equation}
F(\Theta)=F_1(\theta_1)+F_2(\theta_2),
\end{equation}
where \(F_{1(2)}(\theta_{1(2)})\) is the Ginzburg-Landau free energy
for first (second) wire and a simplified notation has been used
\(\theta_1 \equiv \theta_{1,L\leftarrow R}\) and \(\theta_2 \equiv
\theta_{2,L \leftarrow R}\). \(\theta_1\) and \(\theta_2\) are related
to each other and to \(\Theta\) through the phase
constraint~\Eqn{eq:phaseconstraint}.

\begin{widetext}
\subsection{Short nanowires: Josephson junction limit}
If the nanowires are sufficiently short, they may be treated as
Josephson junctions. Unlike the case of long nanowires, described in
the following subsection, in this Josephson regime there is no
metastability, i.e., the free energy of each junction is a
single-valued function of the phase difference, modulo \(2 \pi\),
across it. The phase constraint then implies that there is a rigid
difference between the phases across the two junctions. As a
consequence, \(n_v\) can be set to zero.  The Gibbs free energy in
such a configuration is then
\begin{equation}
\begin{split}
G(I)=-\frac{\hbar}{2 e} \Big(I_{\text{c1}} \cos(\theta_1) +
I_{\text{c2}} \cos(\theta_2) + I \Theta \Big),
\end{split}
\end{equation}
where \(I_{\text{c1}}\) and \(I_{\text{c2}}\) are the critical
currents for the junctions.  In thermodynamic equilibrium, the Gibbs
free energy must be minimized, so the dependent variable \(\Theta\)
must be chosen such that \(\partial G(I)/\partial \Theta=0\).

Using \(\theta_1=\Theta+\delta\) and \(\theta_2=\Theta-\delta\),
\(G(I)\) may be rewritten in the form
\begin{equation}
\begin{split}
\tilde{G}(I)=-\frac{\hbar}{2 e}\Big(
\sqrt{(I_{\text{c1}}+I_{\text{c2}})^2 \cos^2
\delta+(I_{\text{c1}}-I_{\text{c2}})^2 \sin^2 \delta} \cdot
\cos(\vartheta)+I\,\vartheta_1\Big),
\end{split}
\label{GJJ}
\end{equation}
where we have shifted the free energy by an additive constant
\(\left(\frac{\hbar}{2 e}\right) I
\tan^{-1}\left[\left(\frac{I_{\text{c1}}-I_{\text{c2}}}{I_{\text{c2}}
+ I_{\text{c1}}}\right)\tan\delta\right]\), and \(\vartheta
\equiv\Theta+\tan^{-1} \left[ \left(\frac{I_{\text{c1}} -
I_{\text{c2}}}{I_{\text{c2}} + I_{\text{c1}}} \right)
\tan\delta\right]\).  In this model, the option for having
\(I_{\text{c1}}\neq I_{\text{c2}}\) is kept open. Equation~(\ref{GJJ})
shows that, up to an additive constant, the free energy of the
two-junction device is identical to that of an effective
single-junction device with an effective \(I_\text{c}\), which is
given by
\begin{equation}
\label{eq:r1}
I_\text{c}=\sqrt{(I_{\text{c1}}+I_{\text{c2}})^2 \cos^2
\delta+(I_{\text{c1}}-I_{\text{c2}})^2 \sin^2 \delta}.
\end{equation}
Thus, we may determine the resistance of the two-junction device by
applying the well-known results for a single junction, established by
Ivanchenko and Zil'berman~\cite{IZ} and by Ambegaokar and
Halperin~\cite{AH}:
\begin{subequations}
\label{eq:r2}
\begin{align}
R&=R_\text{n} \frac{2(1-x^2)^{1/2}}{x} \exp\left(-\gamma(\sqrt{1-x^2}
+ x \sin^{-1} x)\right) \sinh(\pi \gamma x/2) \label{RJ1}, \\ x&\equiv
I/I_\text{c}, \hskip2cm \gamma\equiv\hbar I_\text{c}/e k_\text{B} T,
\label{RJ3}
\end{align}
\end{subequations}
where \(R_\text{n}\) is the normal-state resistance of the
two-junction device. This formula for \(R\) holds when the free-energy
barrier is much larger than \(k_\text{B} T\), so that the barriers for
phase slips are high. References~\cite{IZ, AH} provide details on how
to calculate the resistance in the general case of an over damped
junction, which includes that of shallow barriers.
Figure~\ref{fig:mr219} shows the fits to the resistance, computed
using \Eqns{eq:r1}{eq:r2}, as a function of temperature, magnetic
field, and total current for sample 219-4. Observe that both the
field- and the temperature-dependence are in good agreement with
experimental data. In Section~\ref{sec:expamp} we make more precise
contact between theory and experiment, and explain how the data have
been fitted.  We also note that, as it should, our Josephson junction
model exactly coincides with our extension of the LAMH model in the
limit of very short wires and for temperatures for which the
barrier-crossing approximation is valid.

\begin{figure*}
\begin{minipage}[]{0.51cm}
(a)
\vskip4cm
\phantom{ }
\end{minipage}
\begin{minipage}[]{7.51cm}
\includegraphics[width=7.5cm]{\figdir/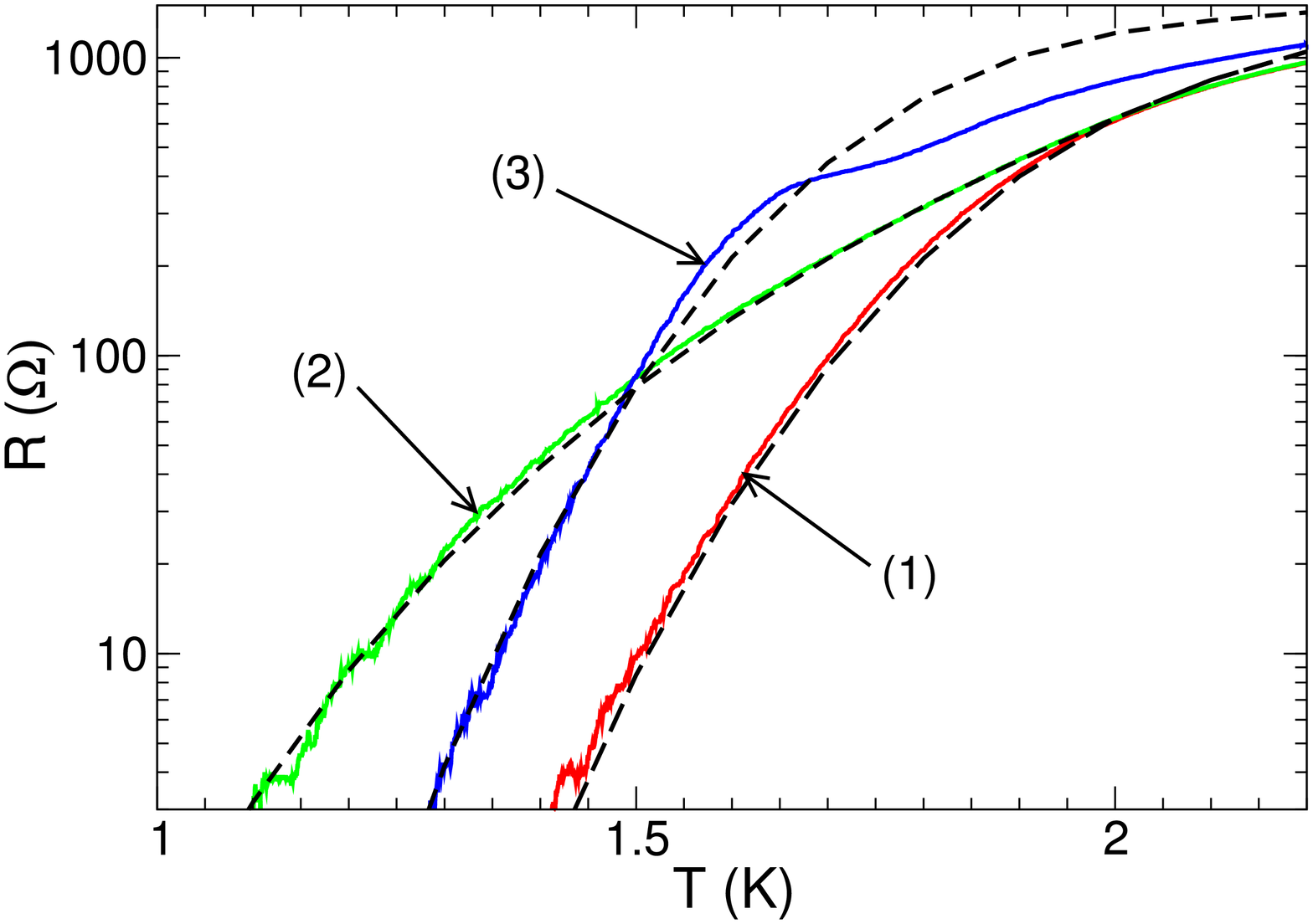}
\end{minipage}
\hskip1cm
\begin{minipage}[]{0.51cm}
(b)
\vskip4cm
\phantom{ }
\end{minipage}
\begin{minipage}{7.51cm}
\includegraphics[width=7.5cm]{\figdir/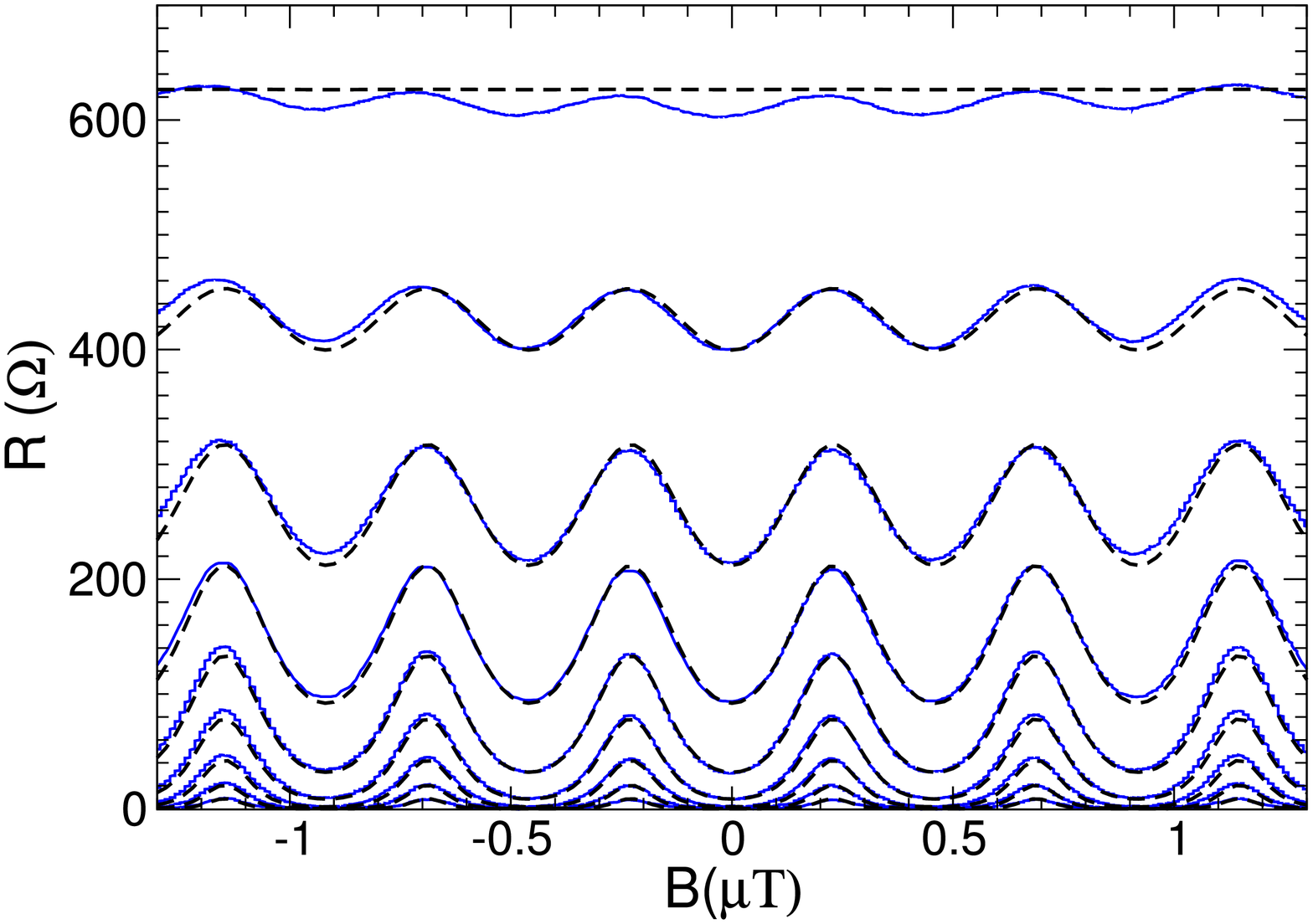}
\end{minipage}
\caption{Sample 219-4: Experimental data (solid lines) and theoretical
fits using the Josephson junction model (dashed lines). (a):
Resistance {\it vs.\/} temperature curves. (1)~Zero magnetic field and
low total current. (2)~Magnetic field set to maximize the resistance
and low total current. (3)~Zero magnetic field and \(70 \, \text{nA}\)
total current.  (b): Resistance as a function of magnetic field at
various temperatures from \(1.2\) to \(2.0\) K in \(0.1\) K
increments. The fitting parameters used were \(J_\text{c1} =
639\,\text{nA}\), \(J_\text{c2} = 330\,\text{nA}\), \(T_\text{c1} =
2.98\,\text{K},\) and \(T_\text{c2} = 2.00\,\text{K}\), with
corresponding coherence lengths \(\xi_1(0) = 23\,\text{nm}\) and
\(\xi_2(0) = 30\,\text{nm}\).  Only one set of fitting parameters
[derived from curves (1) and (2)] was used to produce all the
theoretical curves.}
\label{fig:mr219}
\end{figure*}

\end{widetext}

\subsection{Longer nanowires: LAMH regime}

In this section we describe an extension of the LAMH model of
resistive fluctuations in a single narrow wire~\cite{LA, MH}, which we
shall use to make a quantitative estimate of the voltage across the
two-wire device at a fixed total current.  In this regime the
nanowires are sufficiently long that they behave as LAMH wires.  We
shall only dwell on two-wire systems, but we note in passing that the
model can straightforwardly be extended to more complicated sets of
lead interconnections, including periodic, grating-like arrays (see
Appendix~\ref{app:multi-wire}).

As the sample is not simply connected, i.e.,~there is a hole inside the
AB contour, it is possible that there are multiple metastable states
that can support the total current.  These states differ by the
number of times the phase winds along paths around the AB contour.
The winding number \(n_v\) changes whenever a vortex (or an
anti-vortex) passes across one of the wires.

\begin{figure*}
\includegraphics[width=15cm]{\figdir/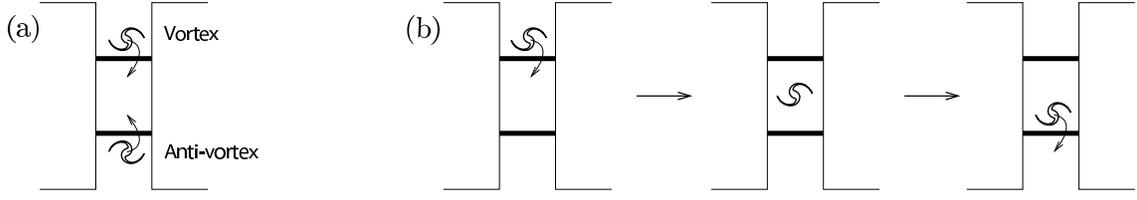}
\caption{Thermally activated phase slip processes under
consideration. (a)~Parallel phase slips. (b)~Sequential phase slips.}
\label{tunnel_pic}
\end{figure*}

In the present theory, we include two kinds of processes that lead to
the generation of a voltage difference between the the leads; see
Fig.~\ref{tunnel_pic}.  In the first kind of process
(Fig.~\ref{tunnel_pic}a), two phase slips occur simultaneously: a
vortex passes across the top wire and, concurrently, an anti-vortex
passes across the bottom wire (in the opposite direction), so that the
winding number remains unchanged.  In the second kind of process
(Fig.~\ref{tunnel_pic}b), the phase slips occur sequentially: a vortex
(or anti-vortex) enters the AB contour by passing across the top (or
bottom) wire, stays inside the contour for some time-interval, and
then leaves the AB contour through the bottom (or top)
wire~\cite{ref:topology_note}.

Our goal is to extend LAMH theory to take into account the influence
of the wires on each other.  In Appendix~\ref{LAMH1}, we review some
necessary ingredients associated with the LAMH theory of a single
wire.  As the wires used in the experiments are relatively short
(i.e.~10 to 20 zero-temperature coherence lengths in length), we also
take care to correctly treat the wires as being of finite length.

Recall that we are considering experiments performed at a fixed total
current, and accordingly, in all configurations of the order parameter
this current must be shared between the top and bottom wires.  We
shall refer to this sharing,
\begin{equation}
I=I_1+I_2,
\end{equation}
as the {\em total current constraint\/}.  Let us begin by considering
a phase-slip event in a device with an isolated wire. While the order
parameter in that wire pinches down, the end-to-end phase accumulation
must adjust to maintain the prescribed value of the current through
the wire. Now consider the two-wire device, and consider a phase slip
event in one of the wires. As in the single-wire case, the phase
accumulation will adjust, but in so doing it will alter the current
flowing through other wire. Thus, in the saddle-point configuration of
the two-wire system the current splitting will differ from that in the
locally stable initial (and final) state.

Taking into account the two kinds of phase-slip processes, and
imposing the appropriate constraints (i.e.~the total current
constraint and the phase constraint), we construct the possible
metastable and saddle-point configurations of the order parameter in
the two-wire system.  Finally, we compute the relevant rates of
thermally activated transitions between these metastable states,
construct a Markov chain~\cite{markov}, and determine the steady-state
populations of these states.  Thus, we are able to evaluate the
time-average of the voltage generated between the leads at fixed
current due to these various dissipative fluctuations.  We mention
that we have not allowed for wires of distinct length or constitution
(so that the Ginzburg-Landau parameters describing them are taken to
be identical).  This is done solely to simplify the analysis;
extensions to more general cases would be straightforward but tedious.

\subsubsection{Parallel pair of nanowires}

The total Gibbs free energy for the two-wire system is given by
\begin{equation}
G(I)=F_1(\theta_{1})+F_2(\theta_2)
-4 \Ecore
\Theta \cdot (J_1+J_2).
\label{G_of_I}
\end{equation}
Here, we have followed MH by rewriting the current-phase term in terms
of dimensionless currents in wires \(i=1,2\), i.e.,~\(J_i\) defined via
\(I_i=8 \pi c J_i \Ecore/\Phi_0\). Moreover, \(\Ecore \equiv
\frac{H_\text{c}^2 \xi \sigma}{8 \pi}\) is the condensate energy density per
unit length of wire, and \(F_i(\theta_i)\) is the Helmholtz free energy
for a single wire along which there is a total phase accumulation of
\(\theta_i\). The precise form of \(F_i(\theta_i)\) depends on whether
the wire is in a metastable or saddle-point state.

We are concerned with making stationary the total Gibbs free energy at
specified total current \(I\), subject to the phase constraint,
\Eqn{eq:phaseconstraint}.  This can be accomplished by making
stationary the Helmholtz free energy on each wire, subject to both the
total current constraint and the phase constraint, but allowing
\(\theta_1\) and \(\theta_2\) to vary so as to satisfy these
constraints---in effect, adopting the total current \(I\) as the
independent variable.  The stationary points of the Helmholtz free
energy for a single wire are reviewed in Appendix~\ref{LAMH1} as
implicit functions of \(\theta_i\), i.e., the end-to-end phase
accumulation along the wire.  The explicit variable used there is
\(J_i\), which is related to \(\theta_i\) via~\Eqn{etoe}.

\subsubsection{Analytical treatment in the limit of long nanowires}

In the long-wire limit, we can compute the resistance analytically by
making use of the single-wire free energy and end-to-end phase
accumulation derived by Langer and Ambegaokar~\cite{LA} (and extended
by McCumber~\cite{M68} for the case of the constant-current
ensemble). Throughout the present subsection we shall be making an
expansion in powers of \(1/b\), where \(b\) is the length of the wire
measured in units of the coherence length, keeping terms only to first
order in \(1/b\).  Thus, one arrives at formul\ae\ for the
end-to-end phase accumulations and Helmholtz free energies for
single-wire metastable ($\text{m}$) and saddle-point ($\text{sp}$)
states~\cite{M68}:
\begin{subequations}
\begin{align}
\theta_{\text{m}}(\kappa)&=\kappa b \label{phi0}, \\
\theta_{\text{sp}}(\kappa)&=\kappa b + 2 \tan^{-1}
\left(\frac{1-3\kappa^2}{2 \kappa^2}\right)^{1/2}, \\
F_{\text{m}}(\kappa)&=-\Ecore \left( b (1-\kappa^2)^2 \right)
\label{F0}, \\ 
F_{\text{sp}}(\kappa)&=-\Ecore \left(b
(1-\kappa^2)^2-\frac{8\sqrt{2}}{3} \sqrt{1-3\kappa^2}\right),
\label{Fsp}
\end{align}
\end{subequations}
where \(\kappa\) is defined via \(J_i=\kappa_i(1-\kappa_i^2)\).  In
the small-current limit, one can make the further simplification that
\(J_i\approx\kappa_i\); henceforth we shall keep terms only up to
first order in \(\kappa\).  To this order, the phase difference along
a wire in a saddle-point state becomes
\begin{equation}
\theta_{\text{sp}}=\kappa b + \pi-2\sqrt{2} \kappa.
\label{phi_sp}
\end{equation}

Next, we make use of these single-wire LAMH results to find the
metastable and saddle-point states of the two-wire system, and use
them to compute the corresponding barrier heights and, hence,
transition rates.  At low temperatures, it is reasonable to expect
that only the lowest few metastable states will be appreciably
occupied.  These metastable states, as well as the saddle-point states
between them, correspond to pairs, \(\kappa_1\) and \(\kappa_2\), one
for each wire, that satisfy the total current constraint as well as
the phase constraint:
\begin{align}
\kappa_1+\kappa_2=J \label{cur5}, \\
\theta_1(\kappa_1)-\theta_2(\kappa_2)=2\pi n_v+2\delta,
\label{cons5}
\end{align}
where we need to substitute the appropriate
\(\theta_{\text{m}/\text{sp}}(\kappa_i)\) from \Eqns{phi0}{phi_sp}
for \(\theta_i(\kappa_i)\).

In the absence of a magnetic field (i.e.~\(\delta=0\)), the lowest
energy state is the one with no circulating current, and the current
split evenly between the two wires. This corresponds to the solution
of \Eqns{cur5}{cons5} with \(n=0\), together with the
substitution~(\ref{phi0}) for \(\theta_i(\kappa_i)\) for both wires
(i.e.~\(\theta_1=\kappa_1 b\) and \(\theta_2=\kappa_2 b\)). Thus we
arrive at the solution:
\begin{subequations}
\begin{align}
\kappa_1&=J/2, & \theta_1&=b J/2, \\
\kappa_2&=J/2, & \theta_2&=b J/2.
\end{align} 
\end{subequations}
If we ignore the lowest (excited) metastable states then only a
parallel phase-slip process is allowed. The saddle point for a
parallel phase slip corresponds to a solution of \Eqns{cur5}{cons5}
with \(n=0\) and the substitution~(\ref{phi_sp}) for
\(\theta_i(\kappa_i)\) for both wires:
\begin{subequations}
\begin{align}
\kappa_1&=J/2, & \theta_1&=b J/2 + \pi-2\sqrt{2} J/2,\\
\kappa_2&=J/2, & \theta_2&=b J/2+ \pi-2\sqrt{2} J/2.
\end{align} 
\end{subequations}
The change in the phase difference across the center of the trench,
\(\Delta \Theta \equiv [\Theta_{\text{sp}}-\Theta_{\text{m}}]\), is
\(\pi-2\sqrt{2}\kappa\) for a forward phase slip, and
\(-\pi-2\sqrt{2}\kappa\) for a reverse phase slip. The Gibbs
free-energy barrier for the two kinds of phase slips, computed by
subtracting the Gibbs free energy for the ground state from that of
the saddle-point state, is
\begin{equation}
\Delta G=\Ecore \left(\frac{16 \sqrt{2}}{3}\pm 4 J \pi\right).
\end{equation}
The former free-energy is obtained by substituting \Eqn{F0} into
\Eqn{G_of_I} for both wires; the latter one is obtained by
substituting \Eqn{Fsp} into \Eqn{G_of_I} for both wires.  We note that
the Gibbs free-energy barrier heights for parallel phase slips (in
both the forward and reverse directions) are just double those of the
LAMH result for a single wire. From the barrier heights, we can work
out the generated voltage by appealing to the Josephson relation, \(V
= (\hbar/2 e) \dot{\Theta}\), and to the fact that each phase
slip corresponds to the addition (or subtraction) of \(2\pi\) to the
phase. Hence, we arrive at the current-voltage relation associated with
parallel phase slips at \(\delta=0\):
\begin{align}
V_{\delta=0 \, \text{, par}}=\frac{\hbar}{e} \, \Omega \, e^{-\beta
\Ecore \frac{16 \sqrt{2}}{3}} \sinh \left(I/I_0\right),
\label{v:d0:par}
\end{align}
where the prefactor \(\Omega\) may be computed using time-dependent
Ginzburg-Landau theory or extracted from experiment, and \(I_0=4
e/\beta h\).

\begin{figure*}
\flushleft{
\begin{minipage}{13.01cm}
\includegraphics[width=13cm]{\figdir/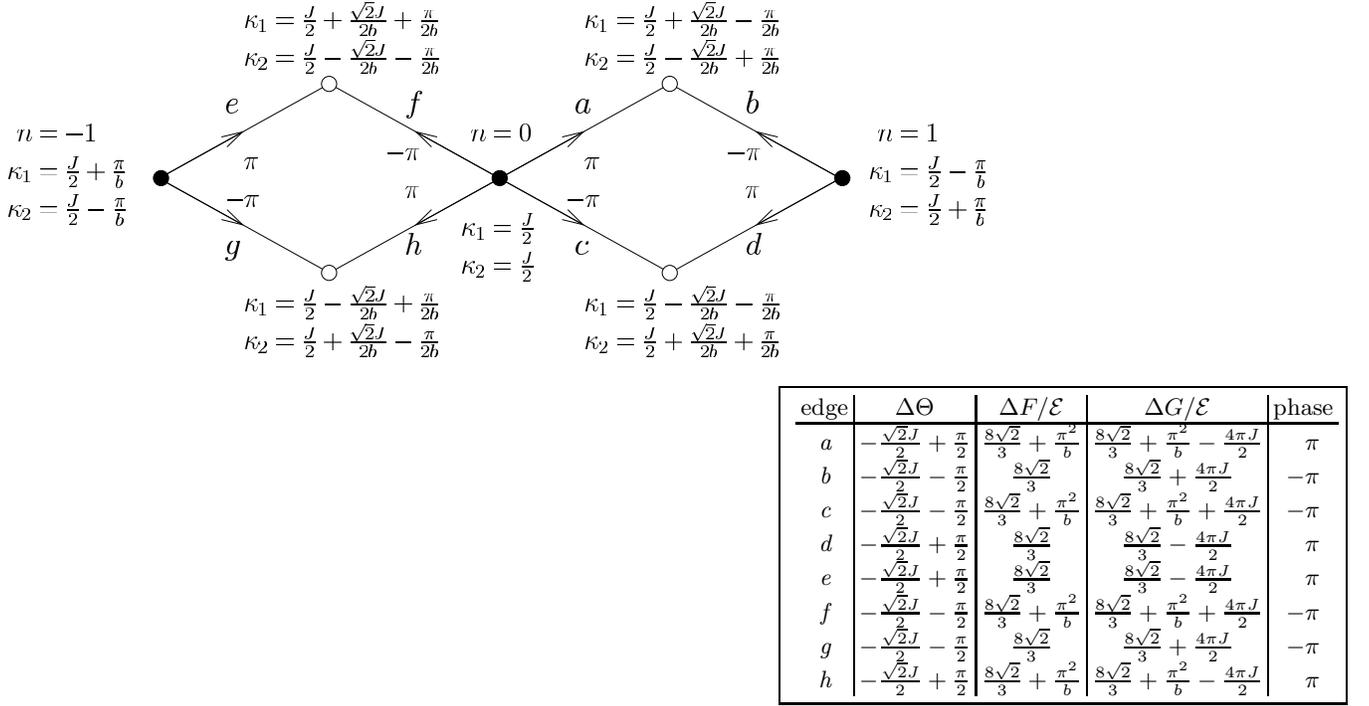}
\end{minipage}
}

\flushright{
\fbox{
\begin{tabular}{c | c | c | c | c}
edge & \(\Delta \Theta\) & \(\Delta F/\Ecore\) & \(\Delta G/\Ecore\) & phase \\
\hline
{\it a} & \(-\frac{\sqrt{2} J}{2}+\frac{\pi}{2}\) & 
\(\frac{8\sqrt{2}}{3}+\frac{\pi^2}{b}\) &
\(\frac{8\sqrt{2}}{3}+\frac{\pi^2}{b}-\frac{4 \pi J}{2}\)& 
\(\phantom{-}\pi\) \\
{\it b} & \(-\frac{\sqrt{2} J}{2}-\frac{\pi}{2}\) & 
\(\frac{8\sqrt{2}}{3}\) &
\(\frac{8\sqrt{2}}{3}+\frac{4 \pi J}{2}\)&
\(-\pi\) \\
{\it c} & \(-\frac{\sqrt{2} J}{2}-\frac{\pi}{2}\) & 
\(\frac{8\sqrt{2}}{3}+\frac{\pi^2}{b}\) &
\(\frac{8\sqrt{2}}{3}+\frac{\pi^2}{b}+\frac{4 \pi J}{2}\)&
\(-\pi\) \\
{\it d} & \(-\frac{\sqrt{2} J}{2}+\frac{\pi}{2}\) & 
\(\frac{8\sqrt{2}}{3}\) &
\(\frac{8\sqrt{2}}{3}-\frac{4 \pi J}{2}\)&
\(\phantom{-}\pi\) \\
{\it e} & \(-\frac{\sqrt{2} J}{2}+\frac{\pi}{2}\) & 
\(\frac{8\sqrt{2}}{3}\) &
\(\frac{8\sqrt{2}}{3}-\frac{4 \pi J}{2}\)&
\(\phantom{-}\pi\) \\
{\it f} & \(-\frac{\sqrt{2} J}{2}-\frac{\pi}{2}\) & 
\(\frac{8\sqrt{2}}{3}+\frac{\pi^2}{b}\) &
\(\frac{8\sqrt{2}}{3}+\frac{\pi^2}{b}+\frac{4 \pi J}{2}\)& 
\(-\pi\) \\
{\it g} & \(-\frac{\sqrt{2} J}{2}-\frac{\pi}{2}\) & 
\(\frac{8\sqrt{2}}{3}\) &
\(\frac{8\sqrt{2}}{3}+\frac{4 \pi J}{2}\)&
\(-\pi\) \\
{\it h} & \(-\frac{\sqrt{2} J}{2}+\frac{\pi}{2}\) & 
\(\frac{8\sqrt{2}}{3}+\frac{\pi^2}{b}\) &
\(\frac{8\sqrt{2}}{3}+\frac{\pi^2}{b}-\frac{4 \pi J}{2}\)&
\(\phantom{-}\pi\) 
\end{tabular}}
}
\caption{Diagram representing the ground state (central filled
circle), the two lowest-energy metastable states (left and right
filled circles), and the saddle-point states connecting them (open
circles) for the case of magnetoresistance minimum
(i.e.~\(\delta=0,\pi,\ldots\)).  The saddle-point states at the top of
the graph correspond to phase slips in the top wire (i.e.~wire \(1\));
those at the bottom correspond to the bottom wire (i.e.~wire
\(2\)). (Saddle-point states for parallel phase slips are not
shown.)\thinspace\ For each state \(\kappa_1\) and \(\kappa_2\) are
listed.  For each barrier (represented by an edge and labeled by {\em
a\/} through {\em h\/}), the table at the right lists the gain in
phase across the trench, the gain in Helmholtz free energy, the
barrier height (i.e.~the gain in Gibbs free energy), and the amount of
phase that would effectively be generated at the end of the phase slip
event (i.e.~upon completion of a closed loop, the amount of phase
generated is the sum of the effective phases). }
\label{0_diag}
\end{figure*}

If we take into account the two lowest excited states, which we
ignored earlier, then voltage can also be generated via sequential
phase slips (in addition to the parallel ones, treated above). To
tackle this case, we construct a diagram in which the vertices
represent the metastable and saddle-point solutions of
\Eqns{cur5}{cons5}, and the edges represent the corresponding free
energy barriers; see Fig.~\ref{0_diag}. Pairs of metastable-state
vertices are connected via two saddle-point-state vertices,
corresponding to a phase slip on either the top or the bottom wire. To
go from one metastable state to another, the system must follow the
edge out of the starting metastable state leading to the desired
saddle-point state. We assume that, once the saddle-point state is
reached, the top of the barrier has been passed and the order
parameter relaxes to the target metastable state. (To make the graph
more legible, we have omitted drawing the edge that corresponds to this
relaxation process.)  To find the Gibbs free-energy difference between
a metastable state and a saddle-point state, we need to know the phase
difference across the center of the trench. To resolve the ambiguity
of \(2 \pi\) in the definition of \(\Theta\), the phase difference can
be found by following the wire with no phase slip.  To further improve
the legibility of Fig.~\ref{0_diag}, the free-energy barriers are
listed in a separate table to the right. Note, that a phase slip on
just one of the wires, being only half of the complete process, can be
regarded to a gain in phase of \(\pm \pi\) for the purposes of
calculating voltage, as indicated in both the graph and the table.

Once the table of barrier heights has been computed, we can construct
a Markov chain on a directed graph, where the metastable states are
the vertices---in effect, an explicit version of our diagram. In
general, each pair of neighboring metastable states, \(s_{n}\) and
\(s_{n+1}\), are connected by four directed edges:
\begin{subequations}
\begin{align}
s_{n}&\xrightarrow[\text{top}]{\hskip1.1cm} s_{n+1} &
s_{n}&\xrightarrow[\text{bottom}]{\hskip1.1cm} s_{n+1}\\ 
s_{n}&\xleftarrow[\text{top}]{\hskip1.1cm} s_{n+1} &
s_{n}&\xleftarrow[\text{bottom}]{\hskip1.1cm} s_{n+1}
\end{align}
\end{subequations}
where the probability to pass along a particular edge is given by
\(P(\cdot)=\exp{-\beta \Delta G_{(\cdot)}}\), in which \(\Delta
G_{(\cdot)}\) may be read off from the table in Fig.~\ref{0_diag}.

We denote the occupation probability of the \(n^\text{th}\) metastable
state by \(o_n\), where \(n\) corresponds to the \(n\) in the phase
constraint~(\ref{eq:phaseconstraint}). \(o_n\) may be computed in the
standard way, by diagonalizing the matrix representing the Markov
chain~\cite{markov}. Each move in the Markov chain can be associated
with a gain in phase across the device of \(\pm \pi\), as specified in
Fig.~\ref{0_diag}. Thus, we may compute the rate of phase-gain, and
hence the voltage:
\begin{equation}
V=\frac{\Omega \hbar}{4 e} \sum_{\langle n m \rangle} 
\frac{o_n}{g_{n,m}}
\big(P(s_{n}\xrightarrow[\text{top}]{} s_{m})
-P(s_{n}\xrightarrow[\text{bot}]{} s_{m})\big),
\end{equation}
where the rate prefactor \(\Omega\) is to be determined, \(\langle n m
\rangle\) indicates that the sum runs over neighboring states only,
and \(g_{n,m}\) keeps track of the sign of the phase-gain for reverse
phase-slips:
\begin{equation}
g_{n,m}=\left\{
\begin{array}{rc}
 1,& \text{ if }m>n, \\
-1,& \text{ if }m<n. 
\end{array}
\right.
\end{equation}
For the case \(\delta=0\), and keeping the bottom three states only, the
voltage generated via sequential phase slips turns out to be
\begin{align}
V_{\delta=0 \, \text{, seq}}=\frac{2 \hbar}{e} \,\Omega \, e^{-\beta \Ecore
\left(\frac{8 \sqrt{2}}{3}+\frac{\pi^2}{b}\right)} \sinh(I/2 I_0).
\label{v:d0:seq}
\end{align}

Having dealt with the case of \(\delta=0\) (and hence obtained the
value of the resistance at magnetic fields corresponding to resistance
minima), we now turn to the case of \(\delta=\pi/2\), i.e., resistance
maxima.

\begin{figure*}
\begin{minipage}{8.1cm}
\includegraphics[width=8cm]{\figdir/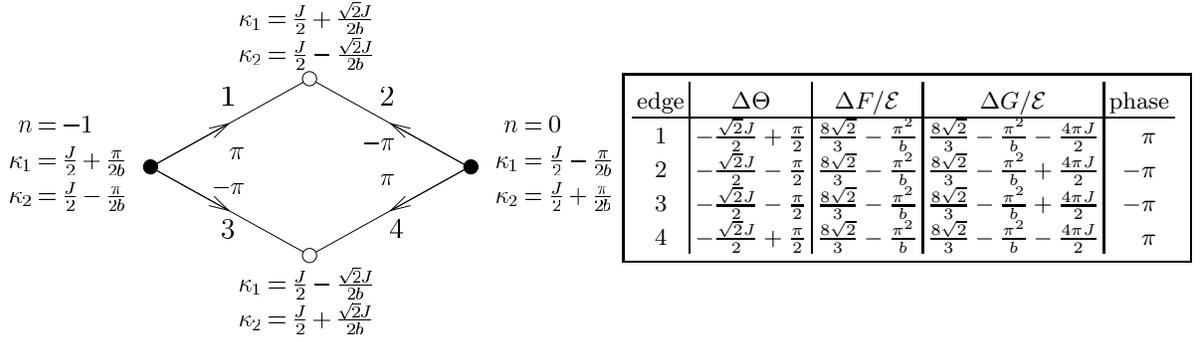}
\end{minipage}
\fbox{\begin{tabular}{c|c|c|c|c}
edge & \(\Delta \Theta\) & \(\Delta F/\Ecore\) & \(\Delta G/\Ecore\) & phase \\
\hline
1 & \(-\frac{\sqrt{2} J}{2}+\frac{\pi}{2}\) & 
\(\frac{8\sqrt{2}}{3}-\frac{\pi^2}{b}\) &
\(\frac{8\sqrt{2}}{3}-\frac{\pi^2}{b}-\frac{4 \pi J}{2}\)& 
\(\phantom{-}\pi\) \\
2 & \(-\frac{\sqrt{2} J}{2}-\frac{\pi}{2}\) & 
\(\frac{8\sqrt{2}}{3}-\frac{\pi^2}{b}\) &
\(\frac{8\sqrt{2}}{3}-\frac{\pi^2}{b}+\frac{4 \pi J}{2}\)&
\(-\pi\) \\
3 & \(-\frac{\sqrt{2} J}{2}-\frac{\pi}{2}\) & 
\(\frac{8\sqrt{2}}{3}-\frac{\pi^2}{b}\) &
\(\frac{8\sqrt{2}}{3}-\frac{\pi^2}{b}+\frac{4 \pi J}{2}\)&
\(-\pi\) \\
4 & \(-\frac{\sqrt{2} J}{2}+\frac{\pi}{2}\) & 
\(\frac{8\sqrt{2}}{3}-\frac{\pi^2}{b}\) &
\(\frac{8\sqrt{2}}{3}-\frac{\pi^2}{b}-\frac{4 \pi J}{2}\)&
\(\phantom{-}\pi\) 
\end{tabular}
}
\caption{Diagram and corresponding Table for the case of
magnetoresistance maxima (i.e.~\(\delta=\pi/2,3 \pi/2,\ldots\)).  See
the caption to Fig.~\ref{0_diag} for explanation of the diagram.}
\label{piby2_diag}
\end{figure*}

In this half-flux quantum situation, there are two degenerate
lowest-energy states, with opposite circulating currents.  These
states are connected by saddle-point states in which a phase-slip is
occurring on either the top or bottom wire. The diagram of the
degenerate ground states and the saddle-point states connecting them
is shown in Fig.~\ref{piby2_diag}. By comparing the diagram with the
associated Table, it is easy to see that the free-energy barriers are
biased by the current, making clockwise traversals of
Fig.~\ref{piby2_diag} more probable than counter-clockwise
traversals. As there are only two metastable states being considered,
and as they are degenerate, it is unnecessary to go through the Markov
chain calculation; clearly, the two states each have a population of
\(1/2\). The voltage being generated by the sequential phase-slip is
then given by
\begin{align}
V_{\delta=\pi/2 \text{, seq}}=\frac{\hbar}{2 e} \, \Omega \, e^{-\beta
\Ecore \left( \frac{8 \sqrt{2}}{3}-\frac{\pi^2}{b}\right)} \sinh(I/2 I_0).
\label{v:dpb2:seq}
\end{align}
\(V_{\delta=\pi/2 \text{, seq}}\) is larger than the sum of
\(V_{\delta=0 \text{, seq}}\) and \(V_{\delta=0 \text{, par}}\), so,
as expected, the resistance is highest at magnetic fields
corresponding to \(\delta=\pi/2\).  For very long wires, the
perturbation of one wire when a phase slip occurs in the other is very
small, and therefore we expect that the dependence of resistance on
magnetic field will decrease with wire length. Indeed, for very long
wires, the difference in barrier heights to sequential phase slips
between the \(\delta=0\) and \(\delta=\pi/2\) cases disappears
(i.e.~\Eqn{v:d0:seq} and \Eqn{v:dpb2:seq} agree when \(b \gg 1\)).

\subsubsection{Numerical treatment for intermediate-length nanowires}

\begin{figure}
\includegraphics[]{\figdir/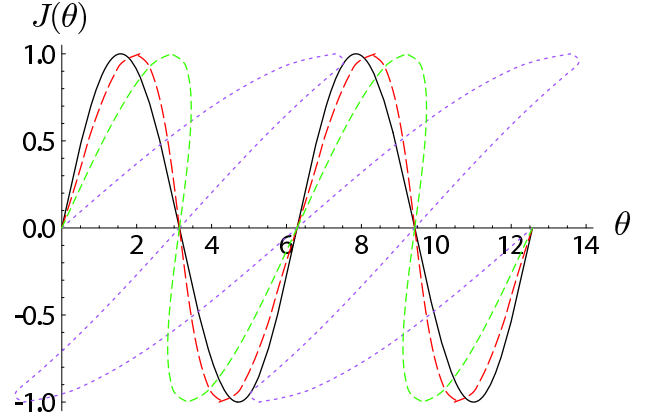}
\caption{\label{fig:jf_vs_t} Current (in units of the critical
current) vs.~end-to-end phase accumulation for superconducting wires
of various lengths: \(0\xi\) (solid line), \(1.88\xi\), \(5.96\xi\),
\(14.4\xi\) (dotted line). The transition from LAMH to Josephson
junction behavior is evident from the loss of multivaluedness of the
current, as the wire length is reduced.}
\end{figure}

Instead of using the long-wire approximation, \Eqnsd{phi0}{Fsp}, we
can use the exact functions for the end-to-end phase accumulation
along a wire \(\theta(J(\kappa))\), and the Helmholtz free energy
\(F_{\text{m}/\text{sp}}(J(\kappa))\). By dropping the long-wire
approximation, as the temperature approaches \(T_\text{c}\) and the
coherence length decreases the picture correctly passes to the
Josephson limit.  In this approach, the total current and the phase
constraints must be solved numerically, as \(\theta(J(\kappa))\) is a
relatively complicated function. Figure~\ref{fig:jf_vs_t} provides an
illustration of how, for a single wire, the function \(J(\theta)\)
depends on its length.  We shall, however, continue to use the
barrier-crossing approximation. Because the barriers get shallower
near \(T_\text{c}\), our results will become unreliable (and, indeed,
incorrect) there.

\begin{figure}
\includegraphics[width=8cm]{\figdir/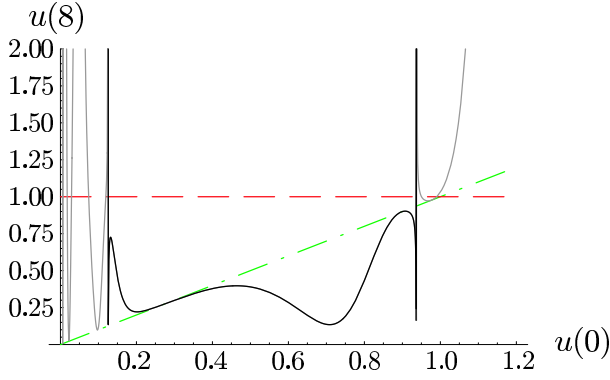}
\caption{Squared amplitude \(u(b/2)\) of the order parameter at the
end of a wire, as a function of its value \(u_0=u(0)\) at the
mid-point of the of the wire, computed using the \(\text{JacobiSN}\)
function [see \Eqn{f_of_z}], for the case \(b=16\). The black line
corresponds to trajectories that do not go through a pole; the gray
line corresponds to trajectories that do pass through at least one
pole. The intersection of the dashed and black lines represents those
trajectories that satisfy the boundary condition \(u(\pm b/2)=1\).
(The intersection of the dashed-dotted and black lines represent
trajectories that start and stop at the same point,
i.e., \(u(b/2)=u_0\)).}
\label{u_of_u_0}
\end{figure}

The form of the order parameter that satisfies the Ginzburg-Landau
equation inside the wire is expressed in \Eqns{f_of_z}{cl}.
Therefore, to construct the functions \(\theta(J)\) and
\(F_{\text{m}/\text{sp}}(J)\) [i.e. \Eqns{etoe}{fexact}], we need to
find \(u_0(J)\), i.e., the squared amplitude of the order parameter in
the middle of the wire. Hence, we need to ascertain suitable boundary
conditions obeyed by the order parameter at the ends of the wire.  For
thin wires, a reasonable hypothesis is that the amplitude of the order
parameter at the ends of the wire matches the amplitude in the leads:
\begin{equation}
f(z=\pm b/2)^2=\frac{H_{\text{c}\,\,\text{film}}^2(T) \,
\xi_\text{film}^2(T)}{H_{\text{c}\,\,\text{wire}}^2(T) \,
\xi_\text{wire}^2(T)}.
\end{equation}
For wires made out of superconducting material the same as (or weaker
than) the leads, this ratio is always larger than
unity~\footnote{In finding \(u_0\) there is a minor numerical
difficulty. As the amplitude of the order parameter is expressed via
the \(\text{JacobiSn}\) function, and \(\text{JacobiSn}[z
\sqrt{u_2/2},u_1/u_2]\) is a doubly periodic function in the first
variable, it is not obvious whether \(\pm (b/2) \sqrt{u_2/2}\) lies in
the first period, as can be seen from Fig.~\ref{u_of_u_0}. As the
trajectory must be simply periodic, \(z \sqrt{u_2/2}\) must intersect
either a zero or a pole in the first unit quarter cell of the
\(\text{JacobiSn}\) function. Now, we are only interested in
trajectories that escape to \(f\rightarrow \infty\) [as \(f(\pm b/2)\)
is assumed to be greater than or equal to unity], so a pole must be
intersected. (However, being outside the first period is unphysical,
as it means that somewhere along the wire \(f=\infty\).) There are
exactly two poles in the first unit quarter cell. They are located at
\(2 v_1+v_2\) and \(v_2\), where \(v_1 \equiv {\rm K}(u_1/u_2)\) and
\(v_2 \equiv i {\rm K}(1-u_1/u_2)\), in which \({\rm K}(\cdot)\) is
the complete elliptic integral. So, instead of checking whether \(\pm
(b/2) \sqrt{u_2/2}\) is outside the unit quarter cell, we can just
determine which pole \(z \sqrt{u_2/2}\) intersects and then see if
\(\pm (b/2) \sqrt{u_2/2}\) lies beyond that pole or not.}  .

Once we have computed the functions \(\theta(J)\) and
\(F_{\text{m}/\text{sp}}(J)\) for both saddle-point and metastable
states on a single wire, we can use the phase and total current
constraints to build the saddle-point and metastable states for the
two-wire device. We proceed as before, by constructing a Markov chain
for the state of the device, except that now we include in the graph
all metastable states of the device. By diagonalizing the Markov
chain, we find the populations of the various metastable states and,
hence, the rate of gain of \(\Theta\).

We plot the typical magnetoresistance curves for various temperatures,
obtained numerically, as well as \(dV/dI\) {\it vs.\/} \(T\) for
various magnetic fields and total currents.  Notice that the
resistance at \(\delta=0\) and large total currents can exceed that at
\(\delta=\pi/2\) with low total-current.

\section{Connections with experiment}
\label{sec:experiment}
In this section a connection is made between our calculations and our
experiments~\cite{us_in_science}. First, the predicted period of the
magnetoresistance oscillations is compared to the experimentally
obtained one. Then, the experimentally-obtained resistance
vs.~temperature curves are fitted using our extension of the IZAH
Josephson junction model (for shorter wires) and our extension of the
LAMH wire model (for longer wires).

\subsection{Device fabrication}
Four different devices were successfully fabricated and measured.  The
devices were fabricated by suspending DNA molecules across a trench
and then sputter coating them with the superconducting alloy of
MoGe. The leads were formed in the same sputter-coating step, ensuring
seamless contact between leads and the wires.  Next, the leads were
truncated lithographically to the desired width.  In the case of
device 930-1, after being measured once, its leads were further
narrowed using focused ion beam milling, and the device was
remeasured.  For further details of the experimental procedure see
Ref.~\cite{us_in_science}.

\subsection{Comparison between theory and experiment}

\subsubsection{Oscillation period}
The magnetoresistance periods obtained for four different samples are
summarized in Table~\ref{table:period}. The corresponding theoretical
periods were calculated using \Eqn{p1}, based on the geometry of the
samples which was obtained via scanning electron microscopy.  To test
the theoretical model, the leads of one sample, sample 930-1, were
narrowed using a focused ion beam mill, and the magnetoresistance of
the sample was remeasured.  The theoretically predicted periods all
coincide quite well with the measured values, except for sample 219-4,
which was found to have a ``+'' shaped notch in one of the leads
(which was not accounted for in calculating the period). The notch
effectively makes that lead significantly narrower, thus increasing
the magnetoresistance period, and this qualitatively accounts for the
discrepancy.

For all samples, when the leads are driven into the vortex state, the
magnetoresistance period becomes much longer, approaching the
Aharonov-Bohm value for high fields.  This is consistent with the
theoretical prediction that the period is then given by \Eqn{p1}, but
with \(l\) replaced by the field-dependent inter-vortex spacing \(r\).

\begin{table*}
\begin{tabular}{l c rrl rrl rr@{.}l rr@{.}l r}
Sample & 
$b \, (\text{nm})$ & \multicolumn{3}{c}{$2a \, (\text{nm})$} &  \multicolumn{3}{c}{$2l \,(\text{nm})$} & 
\multicolumn{3}{c}{Theoretical period \((\mu\text{T})\)} & \multicolumn{3}{c}{Measured period} \((\mu\text{T})\) & error \\ 
\hline
205-4          & $121$ && $266$&&& $11267$ &&& $929$&$21$ &\phantom{hkj}& $947$&$$   & $1.9\%$ \\ 
219-4          & $137$ && $594$&&& $12062$ && \phantom{hkjdad} & $388$&$73$ && $456$&$6$ & $12.8\%$ \\ 
930-1          & $141$ && $2453$&&& $14480$ &&& $78$&$41$ && $77$&$5$  & $-1.2\%$ \\ 
930-1 (shaved) & $141$ && $2453$&&& $8930$  &&& $127$&$14$ && $128$&$3$ & $0.9\%$ \\
205-2          & $134$ && $4046$&&& $14521$ &&& $47$&$41$ && $48$&$9$  & $3.0\%$ 
\end{tabular}
\caption{Comparison between measured and theoretical magnetoresistance
periods.  The geometries of the samples were obtained via scanning
electron microscopy and used to compute the periods theoretically; see
the text for additional details.}
\label{table:period}
\end{table*}

\subsubsection{Oscillation amplitude}
\label{sec:expamp}
\begin{figure*}
\includegraphics[width=7.5cm]{\figdir/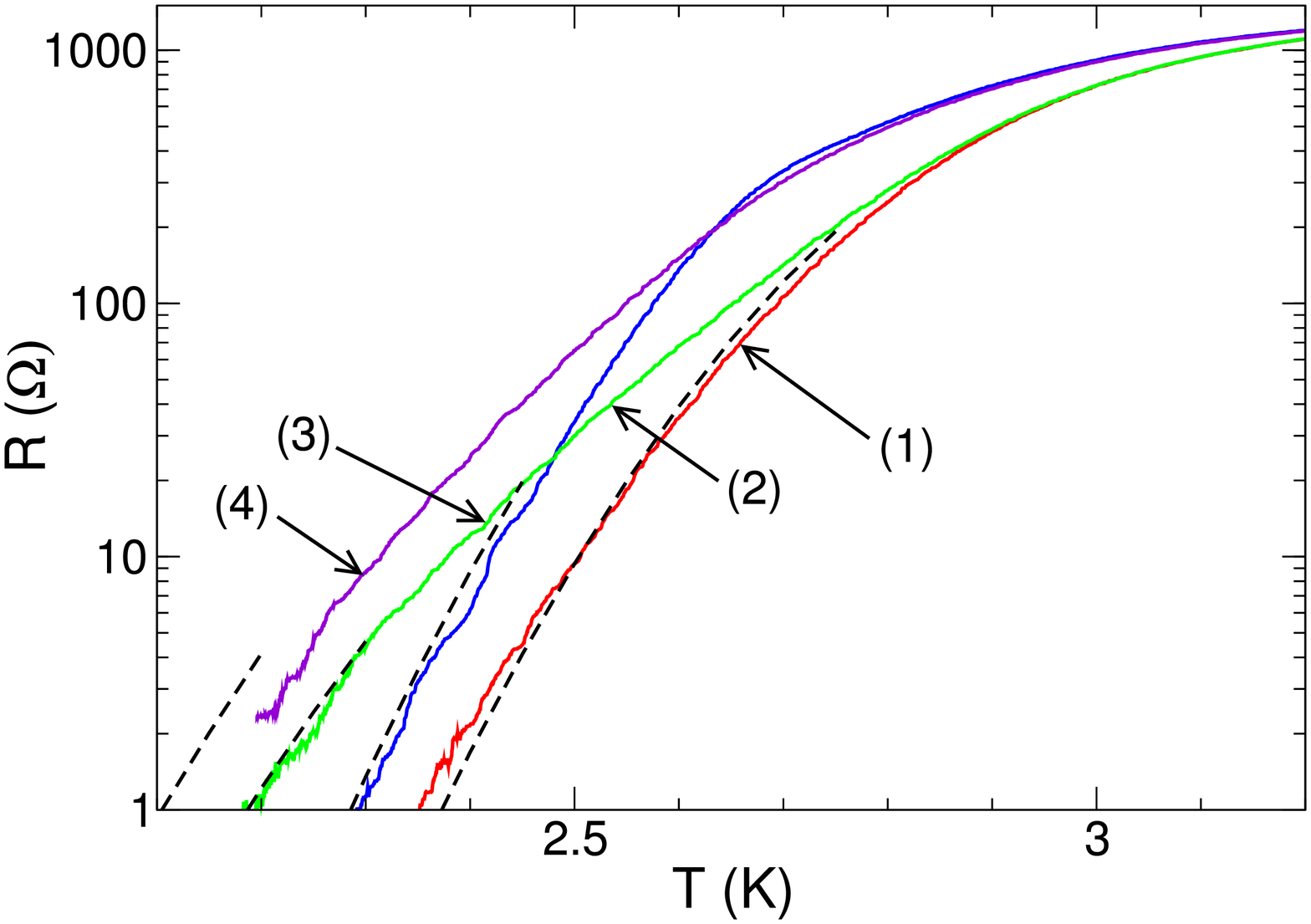}
\hskip1.5cm
\includegraphics[width=7.5cm]{\figdir/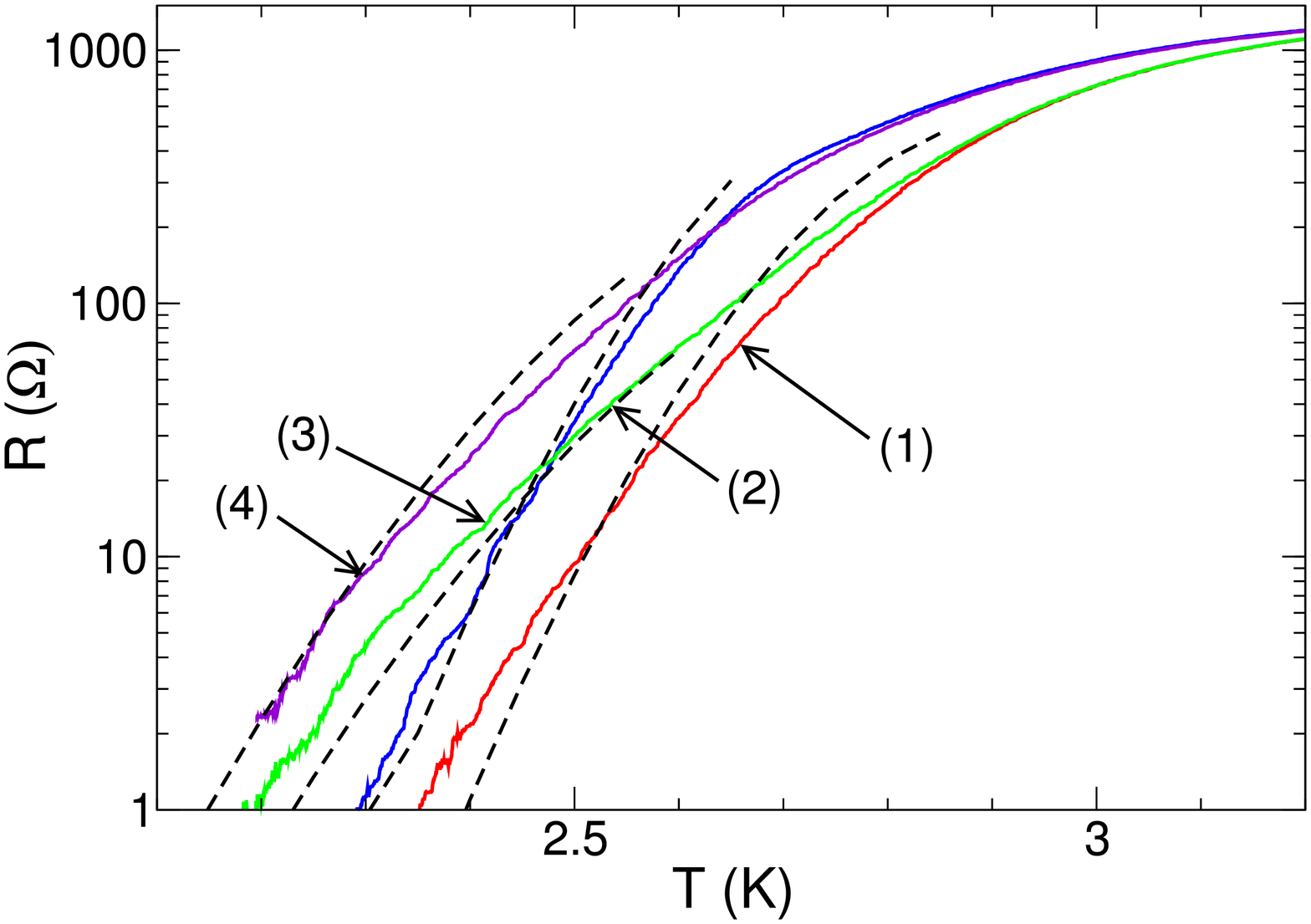}
\begin{tabular}{l c c c c c c}
(LHS) &\(R_1=2882.9 \, \Omega\),     &\(R_2=2941.7\, \Omega\), 
      &\(\xi_{01}=17.3\,\text{nA}\), &\(\xi_{02}=8.7\,\text{nA}\),
      &\(T_{\text{c}1}=3.147\,\text{K}\),      &\(T_{\text{c}2}=3.716\,\text{K}\).\\
(RHS) &\(R_1=2912 \, \Omega\),       &\(R_2=2912 \, \Omega\),
      &\(\xi_{01}=10\,\text{nA}\),   &\(\xi_{02}=9\,\text{nA}\),
      &\(T_{\text{c}1}=3\,\text{K}\),          &\(T_{\text{c}2}=3.65\,\text{K}\).  
\end{tabular}
\caption{Sample 930-1: Resistance vs.~temperature curves. Experimental
data (full lines) and theoretical fits using the LAMH type model in
the intermediate regime (dashed lines).  Theoretical curves terminate
when the short-wire regime is reached, i.e., \(5 \xi(T) \sim
b\). (1)~Zero magnetic field and low total current. (2)~Magnetic field
set to maximize the magnetoresistance and low total current. (3)~Zero
magnetic field and \(80 \, \text{nA}\) total current. (4)~Magnetic
field set to maximize the resistance and \(80 \, \text{nA}\) total
current.  The fit on the left was optimized numerically, and the one
on the right was obtained by hand, showing that a more realistic value
of \(\xi_{01}\) is remains reasonably consistent.  }
\label{fig:mr930}
\end{figure*}

We have made qualitative and quantitative estimates of the resistance
of two-bridge devices in several limiting cases.  For devices
containing extremely short wires [\(b\approx\xi(T)\)], such as sample
219-4, the superconducting wires cannot support multiple metastable
states, and thus they operate essentially in the Josephson junction
limit, but with the junction critical current being a function of
temperature given by LAMH theory as
\(I_\text{c}(T)=I_\text{c}(0)(1-T/T_\text{c})^{3/2}\).  A summary of
fits to the data for this sample, using the Josephson junction limit,
is shown in Fig.~\ref{fig:mr219}.  On the other hand, for longer wires
it is essential to take into account the multiple metastable states,
as is the case for sample 930-1, which has wires of intermediate
length.  A summary of numerical fits for this sample is shown in
Fig.~\ref{fig:mr930}.  In all cases, only the two low total-current
magnetoresistance curves were fitted. By using the extracted fit
parameters, the high total-current magnetoresistance curves were
calculated, with their fit to the data serving as a self-consistency
check. As can be seen from the fits, our model is consistent with the
data over a wide range of temperatures and resistances.  We remark,
however, that the coherence length required to fit the data is
somewhat larger than expected for MoGe.

\section{Concluding remarks}
\label{sec:conclusion}
The behavior of mesoscale NQUIDs composed of two superconducing leads
connected by a pair of superconducting nanowires has been
investigated.  Magnetoresistance measurements~\cite{us_in_science} have
revealed strong oscillations in the resistance as a function of
magnetic field, and these were found to have anomalously short
periods.  The period has been shown to originate in the gradients in
the phase of the superconducting order parameter associated with
screening currents generated by the applied magnetic field.  The
periods for five distinct devices were calculated, based on their
geometry, and were found to fit very well with the experimental
results

The amplitude of the magnetoresistance has been estimated via
extensions, to the setting of parallel superconducting wires, of the
IZAH theory of intrinsic resistive fluctuations in a current-biased
Josephson junction for the case of short wires and the LAMH theory of
intrinsic resistive fluctuations in superconducting wires for pairs of
long wires.  In both cases, to make the extensions, it was necessary
to take into account the inter-wire coupling mediated through the
leads.  For sufficiently long wires, it was found that multiple
metastable states, corresponding to different winding numbers of the
phase of the order parameter around the AB contour, can exist and need
to be considered.  Accurate fits have been made to the resistance
vs. temperature data at various magnetic fields and for several
devices by suitably tuning the critical temperatures, zero-temperature
coherence lengths, and normal-state resistances of the nanowires.

As these device are sensitive to the spatial variations in the phase
of the order parameter in the leads, they may have applications as
superconducting phase gradiometers.  Such applications may include the
sensing of the presence in the leads of vortices or of supercurrents
flowing perpendicular to lead edges.

\bigskip\noindent{\it Acknowledgments\/}:
This work was supported by the U.S.~Department of Energy, Division of
Materials Sciences under Award No.~DEFG02-91ER45439, through the
Frederick Seitz Materials Research Laboratory at the University of
Illinois at Urbana-Champaign. AB and DH would like to also acknowledge
support from the Center for Microanalysis of Material DOE Grant
No.~DEFG02-96ER45439, NSF CAREER Grant No.~DMR 01-34770, and the
A.P.~Sloan Foundation.

\appendix

\begin{widetext}
\section{Multi-wire devices}
\label{app:multi-wire}
In this appendix we give an example of how to extend the theory
Presented in this Paper to the case of devices comprising more than
two wires.  In our example, we consider an array of \(n\) identical
short wires (i.e.~wires in the Josephson junction limit) spaced at
regular intervals. We continue to work at a fixed total current and to
ignore charging effects. The end-to-end phase accumulations along the
wires are related to each other as
\begin{align}
\begin{split}
\theta_2&=\theta_1 + 2 \delta, \\
\theta_3&=\theta_1 + 4 \delta, \\
&\vdots \\
\theta_n&=\theta_1 + 2 (n-1) \delta,
\end{split}
\end{align}
i.e., \(\theta_n-\theta_1=2(n-1)\delta\) (for \(n=2,\ldots,N\)),
where \(\delta\) is the phase accumulation in one of the leads between
each pair of adjacent wires. The Gibbs free energy of the multi-wire
subsystem is given by
\begin{equation}
G(I,\theta_1)=-\frac{h}{2 e} \left( I_\text{c} \sum_{m=1}^n \cos \big(
\theta_1 + 2(m-1) \delta \big) + I \theta_1 \right),
\end{equation}
where \(I\) is the total current and we are assuming that the wires
have identical critical currents. As for the two-junction case, this
junction array, is equivalent to a single effective junction.
Figure~\ref{fig:multi-wire} shows the critical current of this
effective junction as a function of \(\delta\) for devices comprising
2, 5, and 15 wires.  The magnetoresistance of such a device then
follows from IZAH theory, i.e., \Eqn{eq:r2}.

\begin{figure}
\includegraphics[width=8cm]{\figdir/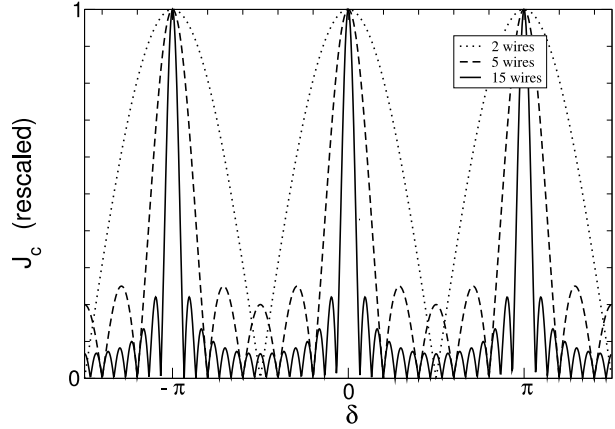}
\caption{Effective single junction critical current for a
multi-junction array, as a function of \(\delta\). The critical current
has been rescaled so that \(J_\text{c}(\delta=0)=1\). Note the similarity
with a multi-slit interference pattern.}
\label{fig:multi-wire}
\end{figure}

\section{Physical Scales}
It is convenient to express the results of the long-wire model,
Eqs.~(\ref{v:d0:par},\ref{v:d0:seq},\ref{v:dpb2:seq}), in terms of
macroscopic physical parameters. Following Tinkham and Lau~\cite{TL},
we express the condensation energy scale per coherence length of wire
as
\begin{equation}
\Ecore=0.22 \, k_\text{B} T_\text{c} \, (1-t)^{3/2} \,
\frac{R_\text{q}}{R_\text{N}}\frac{b}{\xi(T=0)},
\label{RTEcore}
\end{equation}
where \(t\equiv T/T_\text{c}\), \(R_\text{N}\) is the normal-state
resistance of the device, and \(R_\text{q} \equiv h/4e^2 \approx
6.5\,\text{k}\Omega\) is the quantum of resistance.
The LAMH prefactor for sequential phase slips then becomes
\begin{equation}
\Omega=\frac{b \sqrt{1-t}}{\xi(T=0)} \left(\frac{8 \sqrt{2} \,
\Ecore}{3\, k_\text{B} T_\text{c}}\right)^{1/2} \frac{8 k_\text{B}
(T_\text{c}-T)}{\pi \hbar},
\end{equation}
and for parallel phase slips becomes
\begin{equation}
\Omega=\left(\frac{b \sqrt{1-t}}{\xi(T=0)}\right)^2 \left(\frac{16 \sqrt{2} \,
\Ecore}{3\, k_\text{B} T_\text{c}}\right)^{1/2} \frac{8 k_\text{B}
(T_\text{c}-T)}{\pi \hbar}.
\end{equation}
The remaining parameters in the model are \(R_N\), \(T_\text{c}\) and
\(\xi(T=0)\). The normal-state resistance and the critical temperature
may be obtained from the \(R\) vs. \(T\) curve. The coherence
length may be obtained by comparing \(\Ecore\) obtained from the
critical current at low temperature, via
\begin{equation}
I_\text{c}=\frac{2}{3 \sqrt{3}} \frac{16 \pi \Ecore}{\Phi_0},
\end{equation}
with \(\Ecore\) obtained via \Eqn{RTEcore}.

In experiment, it is expected that the two wires are not
identical. The long-wire model can be easily extended to this
case. The number of parameters to be fitted would then expand to
include the normal-state resistance for each wire (only one of which
is free, as the pair are constrained by the normal-state resistance of
the entire device, which can be extracted from the \(R\) vs. \(T\)
curve), a zero-temperature coherence length for each wire, and a
critical temperatures for each wire.

\section{LAMH theory for a single bridge}
\label{LAMH1}

In this appendix we reproduce useful formulas from LA~\cite{LA}, and
rewrite them in a way that is convenient for further calculations,
especially for numerical implementation. As in the case of single-wire
LAMH theory, one starts with the Ginzburg-Landau free energy
\begin{equation}
F=\int_{-b/2}^{b/2} \alpha |\psi|^2+\frac{\beta}{2} |\psi|^4 +
\frac{\hbar^2}{2 m} |\nabla \psi|^2 \, dz.
\end{equation}
The relationships between the parameters of the Ginzburg-Landau free
energy (\(\alpha\) and \(\beta\)), coherence length \(\xi\), the
condensation energy per unit coherence length \(\Ecore\), the critical
field \(H_\text{c}\) and the cross-sectional area of the wire
\(\sigma\) are given by \(\frac{\alpha^2}{\beta}=\frac{H_\text{c}^2
\sigma}{8 \pi}=\Ecore/\xi\) and \(\xi^2=\frac{\hbar^2}{2 m
|\alpha|}\).  Following McCumber~\cite{M68}, it is convenient to work
in terms of the dimensionless units obtained using the
transformations: \(|\psi|^2 \rightarrow \frac{\alpha}{\beta}
|\psi|^2\), \(z \rightarrow \sqrt{\frac{2 m |\alpha|}{\hbar^2}} z\),
and \(b \rightarrow b/\xi=\sqrt{\frac{2 m |\alpha|}{\hbar^2}} b\). In
terms of these units, the free energy becomes
\begin{align}
F= 2 \Ecore \int_{-b/2}^{b/2}
\left(\frac{1}{2}(1-|\psi|^2)^2 +|\nabla \psi|^2 \right) dz.
\end{align}

The Ginzburg-Landau equation is obtained by varying the free energy:
\begin{equation}
\delta F = 0 \,\,\, \Rightarrow \,\,\,
-\psi + |\psi|^2 \psi - \nabla^2 \psi = 0.
\end{equation}
By writing \(\psi=f e^{i \phi}\) and taking the real and imaginary
parts of the Ginzburg-Landau equation one obtains
\begin{align}
-f+f^3+(\phi')^2 f&=f'' \label{feq}, \\
 2 \phi'f'+\phi''f&=0 \label{cl1}.
\end{align}
From \Eqn{cl1}, one finds the current conservation law:
\begin{equation}
f^2 \phi' = J \label{cl},
\end{equation}
where \(J\) is identified with the dimensionless current \(\frac{1}{2
i}(\psi^* \nabla \psi-\psi \nabla \psi^*)\). The physical current (in
stat-amps) is given by \(I=J c H_\text{c}^2 \sigma
\xi/\Phi_0\). Expressing \(\phi'\) in terms of J, \Eqn{feq} becomes
\begin{align}
f''&=-f+f^3+\frac{J^2}{f^3}=-\frac{d}{df}U(f),
\end{align}
where the effective potential \(U(f)\) is given by
\begin{align}
U(f)=\frac{J^2}{2 f^2} +\frac{f^2}{2}-\frac{f^4}{4}.
 \label{pot}
\end{align}
Following LA, Eq.~(\ref{pot}) can usefully be regarded as the equation of
motion for a particle with position \(f(z)\), where \(z\) plays the
role of time, moving in the potential \(U(f)\)~\cite{LA}. Before
proceeding to find the solution of this equation, we pause to consider
the type of trajectories that are possible. Later, it will be
demonstrated that at the edge of the wire \(f(\pm b) \geq 1\), so the
particle starts to the right of the hump; see Fig.~\ref{U_f}. If the
total energy of the particle is less than the height of the hump, the
particle will be reflected by the hump. If, however, the particle starts
with more energy than the height of the hump, it will pass over the
hump and be reflected by the \(J^2/2 f^2\) dominated part of \(U(f)\).

\begin{figure}
\includegraphics{\figdir/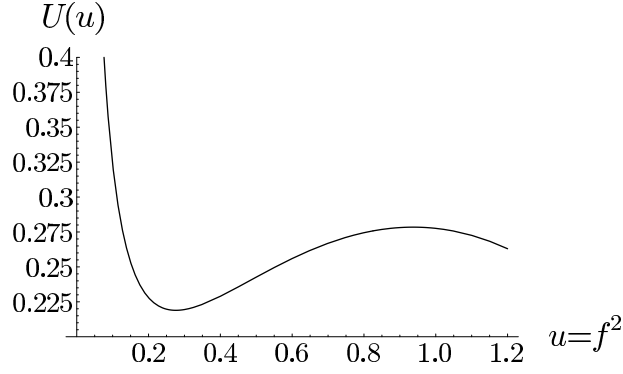}
\caption{``Mechanical potential'' \(U[u=f^2]\) at an intermediate
value of the dimensionless current, plotted as a function
of amplitude squared to make comparison with Fig.~\ref{u_of_u_0} more
convenient.}
\label{U_f}
\end{figure}

The equation of motion can be solved via the first integral
(i.e.~multiplying both sides by \(f\) and integrating with respect to
\(f\)):
\begin{equation}
E=\frac{(f')^2}{2}+U(f)\,\,\,\Rightarrow\,\,\,f'=\sqrt{2(E-U(f))},
\end{equation}
where \(E\) is a constant of integration (i.e.~the energy of the
particle in the mechanical analogy), which gives
\begin{align}
z &= \int_{f_0}^f \frac{df}{\sqrt{2(E-U(f))}}= \int_{f_0}^f \frac{f
df}{\sqrt{2 f^2 E-J^2-f^4+f^6/2}}.
\end{align}
It is convenient to apply ``initial'' conditions at the middle of the
wire, where \(f(z=0)=f_0\), and to integrate towards the edges.  We
require that the particle come back to its starting point after a
``time'' \(b\), i.e.~at the edges of the wire the amplitude of the
order parameter must match the boundary condition. Therefore, the
middle of the wire must be the turning-point for the particle, i.e.,
at \(z=0\) we have \(E=U(f_0)\).

What follows next is a series of manipulations via which one can
express solution for \(f(z)\) in terms of special functions.

Step 1: substitution: \(f^2\rightarrow u\)
\begin{align} 
z&= \frac{1}{2} \int_{u_0}^u
\frac{du}{\sqrt{2 E u-J^2-u^2+u^3/2}}
\end{align} 
Step 2: substitution: \(u\rightarrow u_0+\epsilon\)
\begin{align}
2 z &= \int_{0}^{u-u_0}
\frac{d\epsilon}{
\big[\epsilon \big(\underbrace{\left(\frac{J^2}{u_0}-u_0+u_0^2\right)}_\alpha+\underbrace{\left(\frac{3}{2}
u_0-1\right)}_\beta \epsilon+\epsilon^2/2\big)\big]^{1/2}}
\end{align}
\begin{align} 
2z&=\int_{0}^{u-u_0}
\frac{d\epsilon}{(\epsilon(\epsilon+\beta\epsilon+\epsilon^2/2))^{1/2}}\\
&=\int_{0}^{u-u_0}
\frac{\sqrt{2} d\epsilon}{(\epsilon(\epsilon+\underbrace{\beta
+\sqrt{\beta^2-2\alpha}}_{-u_1})(\epsilon+\underbrace{\beta
-\sqrt{\beta^2-2\alpha}}_{-u_2}))^{1/2}}\\
&=\int_{0}^{u-u_0} \frac{\sqrt{2} d\epsilon}{(\epsilon (\epsilon-u_1)(\epsilon-u_2))^{1/2}}
\end{align}
Step 3: substitution: \(\epsilon\rightarrow u_1 z^2\)
\begin{align}
2 z&=\frac{2\sqrt{2}}{\sqrt{u_2}}\int_{0}^{\sqrt{\frac{u-u_0}{u_1}}} \frac{d \omega}{((\omega^2-1)(\frac{u_1}{u_2}\omega^2-1))^{1/2}}\\
&=\frac{2\sqrt{2}}{\sqrt{u_2}} \text{EllipticF}\big[\text{ArcSin}\big[\sqrt{\frac{u-u_0}{u_1}}\big],\frac{u_1}{u_2}\big] \label{z_of_f}
\end{align}
The following definitions have been used:
\begin{align}
\alpha[u_0] &\equiv J^2/u_0-u_0+u_0^2, & 
\beta[u_0] &\equiv \frac{3}{2} u_0-1,\\
u_1[\alpha,\beta] &\equiv-\beta-\sqrt{\beta^2-2\alpha}, &
u_2[\alpha,\beta] &\equiv-\beta+\sqrt{\beta^2-2\alpha}.
\end{align}

By inverting relation (\ref{z_of_f}) one obtains an explicit equation
for the amplitude of the order parameter as a function of position
along the wire (see Fig.~\ref{u_of_x}):
\begin{subequations}
\begin{align}
f^2(z) &= u_0 + u_1 \sin^2 \big[\text{JacobiAmplitude}\big[z
\sqrt{\frac{u_2}{2}},\frac{u_1}{u_2}\big]\big]\\
&=u_0+u_1 \text{JacobiSn}^2\big[z
\sqrt{\frac{u_2}{2}},\frac{u_1}{u_2}\big] \label{f_of_z}
\end{align}
\end{subequations}

\begin{figure}
\includegraphics[width=7cm]{\figdir/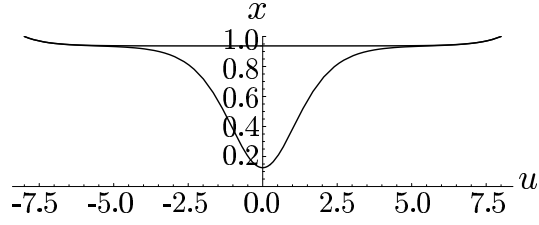}
\caption{Squared amplitude \(u\) of the order parameter as a function
of position along the wire for the two types of solution: metastable
and saddle point.
}
\label{u_of_x}
\end{figure}

The end-to-end phase difference along the wire may be found by 
using the current conservation law. Thus one obtains
\begin{equation}
\theta=\int_{-b/2}^{b/2} \frac{J}{f^2(z)} dz = 2 J \int_0^{b/2} \frac{dz}{
u_0+u_1 \text{JacobiSn}^2\big[z
\sqrt{\frac{u_2}{2}},\frac{u_1}{u_2}\big]}.
\label{etoe}
\end{equation}

The Helmholtz free energy can be found by substituting the expressions
for \(f(z)\) and \(\phi'(z)\) into the expression for the free
energy. One then obtains
\begin{equation}
F=4 \Ecore
\int_0^{b/2} dz (\frac{1}{2}-2 f^2+f^4+J^2/u_0+u_0-u_0^2/2),
\label{fexact}
\end{equation}
where \(E\) was expressed in terms of \(u_0\).  \Eqns{etoe}{fexact}
provide expressions for \(\theta\) and \(F\) which are true regardless
of the length of the wire, and therefore may be used as a starting
point for computing the Gibbs free-energy of the various metastable
states subject to the total-current and the phase constraints.
\end{widetext}


\end{document}